\documentclass[final,5p,twocolumn,authoryear]{elsarticle}

\usepackage{amssymb}
\usepackage{graphicx,amsmath}
\usepackage{ifthen}
\usepackage{xr}
\usepackage{multirow}
\usepackage{url}
\usepackage{hyperref}
\hypersetup{colorlinks=true,citecolor=blue}
\usepackage{breakurl}
\usepackage{endnotes}
\usepackage{float}
\def\beq{\begin{equation}}
\def\eeq{\end{equation}}
\def\bea{\begin{eqnarray}}
\def\eea{\end{eqnarray}}
\def\nn{\nonumber}

\def\degC{$^{\circ}$C}

\biboptions{round,comma,sort&compress}

\journal{Environmental Research Letters}

\begin{document}

\begin{frontmatter}


%\title{Some title needed}
%\title{On global road transport environmental policy making: a comparative distributive data-driven approach}
%\title{Global climate policy-making for private road transport: a comparative distributive data-driven approach}
\title{The effectiveness of policy on consumer choices for private road passenger transport emissions reductions in six major economies}

\author[4cmr]{J.-F. Mercure  \corref{cor1}}
\ead{jm801@cam.ac.uk}

\author[4cmr]{A. Lam}

\address[4cmr]{Cambridge Centre for Climate Change Mitigation Research (4CMR), Department of Land Economy, University of Cambridge, 19 Silver Street, Cambridge, CB3 1EP, United Kingdom}

\cortext[cor1]{Corresponding author: Jean-Fran\c{c}ois Mercure}
\fntext[fn1]{Tel: +44 (0) 1223337126, Fax: +44 (0) 1223337130}

%Note -- 838 words 23/09/2013
\begin{abstract}

The effectiveness of fiscal policy to influence vehicle purchases for emissions reductions in private passenger road transport depends on its ability to incentivise consumers to make choices oriented towards lower emissions vehicles. However, car purchase choices are known to be strongly socially determined, and this sector is highly diverse due to significant socio-economic differences between consumer groups. Here, we present a comprehensive dataset and analysis of the structure of the 2012 private passenger vehicle fleet-years in six major economies across the World (UK, USA, China, India, Japan and Brazil) in terms of price, engine size and emissions distributions. We argue that choices and aggregate elasticities of substitution can be predicted using this data, enabling to evaluate the effectiveness of potential fiscal and technological change policies on fleet-year emissions reductions. We provide tools to do so based on the distributive structure of prices and emissions in segments of a diverse market, both for conventional as well as unconventional engine technologies. We find that markets differ significantly between nations, and that correlations between engine sizes, emissions and prices exist strongly in some markets and not strongly in others. We furthermore find that markets for unconventional engine technologies have patchy coverages of varying levels. These findings are interpreted in terms of policy strategy. 

\end{abstract}


\end{frontmatter}

%\linenumbers


\section{Introduction}

Transport generates 5.3~Gt out of 32.7~Gt of CO$_2$ of global fuel combustion greenhouse gas (GHG) emissions contributing to climate change, and over 30\% of total emissions annual growth \citep{IEAWEB2013}. The current trend in global emissions is leading the world towards global warming exceeding 4\degC, with the most important component emitted by the electricity sector, transport coming second \citep{IPCCAR5WGIIITS2}. Transport is not currently highly regulated for emissions, and consumer preferences with increasing income drives choices towards increasingly carbon intense engines \cite[e.g.][]{Gallachoir2009,Zachariadis2013}.\footnote{According to our data, this is also true in the UK.} The lack of clear policy direction in the sector may be related to a lack of understanding of the impact of policies aiming at incentivising consumer choice \citep{He2011}.

The fuel efficiency of the car fleet depends directly on its composition of engine types and sizes. In other emissions intensive sectors such as industry, technologies of the same type (e.g. boilers, blast furnaces, power plants) differ modestly predominantly due to vintage.\footnote{By vintage we mean different variants of the same technologies of different ages, variations that stem from the gradual incorporation of particular innovations over time.} The transport sector, however, features a very wide continuous array of possible fuel efficiencies (spanning a factor of 3-4, see data below), that do not depend as strongly on the state of technology and age of cars as it does on socio-economic characteristics of owners. As demonstrated by \cite{McShane2012}, vehicle choices strongly relate to social groups, and thus existing socio-economic differences are reflected in the types of vehicles that consumers purchase, consistent with socio-anthropological theory \citep{Douglas1979}. We ask two questions here: do different markets respond differently to policy instruments?  Is it possible to obtain quantitative insight on the effectiveness of proposed policies using market research methods? 

Cars typically survive in the fleet for around 12~years.\footnote{The life expectancy of cars in the UK is 12~years, as calculated by the authors using a survival analysis with data from the \cite{DVLA} data, unpublished.} The carbon intensity of the private vehicle fleet can be reduced by influencing vehicle purchase choices in two different ways:\footnote{Emissions can also be reduced by reductions of car use, for instance with fuel taxes, involving possible transport mode or lifestyle changes, not the subject of this paper.} by improving the average fuel economy of the new fleet-year (i.e. new models) with conventional engine technology, or by gradually replacing conventional combustion engines by hybrid and electric systems. It is clear that comprehensive emissions reduction policy must include coordinated demand-pull and supply-push policies; however, in the short term, it is policies affecting consumer choice that have the most impact, since it takes years to change production lines and supply chains of manufacturers, while sales can change overnight. Here, we focus on demand-pull policies that affect consumer choice.

Potentials for efficiency improvements in transport have been widely explored \citep{IPCCAR5WGIIITS2, WEC2009, IEAETP2012}. In energy-environment policy modelling research, `representative' vehicles are most of the time used, with cost-optimisation frameworks \citep[e.g.][]{McCollum2014,Takeshita2011,vanderZwaan2013}, a procedure that is, as we show here, insufficiently rich in socio-economic patterns, while the optimisation of cost driving agent choice is contradicted by empirical work \citep{McShane2012}. It is also not clear that evaluating the average willingness to pay for higher fuel efficiency is feasible or even insightful in a highly diverse sector where prices span an order of magnitude following the income distribution (see below), and the literature is not unanimous on any value \citep{Greene2010,Anderson2011,Gallagher2011}. A more promising approach appears to be to use marketing research, where the likely response of consumers to policies is inferred from the structure of the market itself \citep[as in][]{He2011}, and this approach naturally introduces socio-economic distributions.

In this paper, we explore the emissions, technical and economical characteristics and potential for emissions reductions of car markets in six major representative economies of the World: the United Kingdom, the United States, China, India, Japan and Brazil. These markets possess very different characteristics that are the result of different histories of policy and regulations, but also different socio-economic characteristics of their populations as well as different cultures. We show that, even with highly globalised car manufacturers, the distribution of car purchases and their fuel efficiencies have radically different characteristics in different markets. We calculate the effectiveness of emissions reduction policies through vehicle choice in terms of emissions reductions per unit tax, applied on the price of vehicles or fuels. This is compared to the effectiveness of policy for technological change. This leads to two considerations: firstly, different car markets may require different policies for effective emissions reductions. Secondly, differences in market characteristics between countries show comparatively what may be realistically achievable elsewhere with policy. Details of calculations are given in the Supplementary Information (SI).

\section{Materials and Methods}

\subsection{Theory for car purchase choice \label{sect:Behaviour}}

\cite{McShane2012} makes a remarkable demonstration, using correlations spatially resolved using postcode data, that vehicle purchases in the USA are visually influenced by previous purchases of similar vehicles locally, but not by purchases that happen at distances where visual influence is weak. This takes place within \emph{geographical areas}, \emph{social identity groups}, and within \emph{vehicle types} and \emph{price tiers}. This work provides strong evidence of how vehicle choices, determined within social groups, take place within restricted subsets the vehicle market in specific price segments. 

\begin{figure}[t]
		\begin{center}
			\includegraphics[width=1\columnwidth]{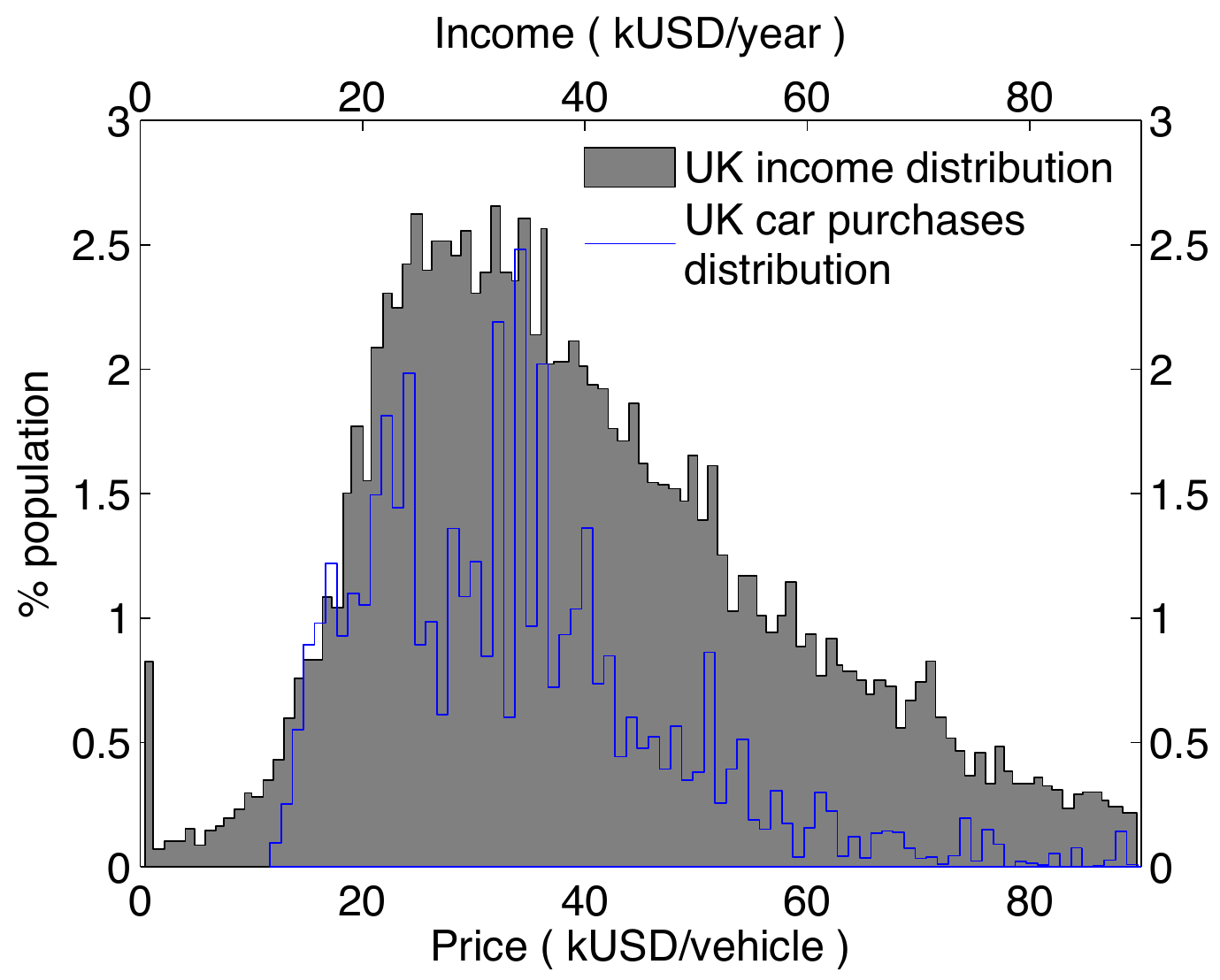}
		\end{center}
	\caption{Comparison between the UK income distribution in kUSD/year after tax \citep[obtained from][]{UKGOV2013} and the distribution of vehicle prices after tax for vehicles purchased in the UK in 2012 (this study, using data from \cite{DVLA} and \cite{CarPages}, see section~\ref{sect:results}). The vehicle price distribution population (the $y$ scaling) is scaled by a per capita vehicle ownership factor of 0.44 (to convert units between per vehicle owner and per capita), which corresponds to about 28.3 million vehicles owned by 63.9 million persons in 2012 \citep{DVLA}. The two distributions are likely connected.}
	\label{fig:UKIncome2012}
\end{figure}

This can be directly observed using socio-economic distribution data. Figure~\ref{fig:UKIncome2012} shows the 2012 distribution of income for the whole UK population \citep[from][]{UKGOV2013}, along with the 2012 UK distribution of car purchase prices (this work, see below) multiplied by average car ownership. Both datasets are well described by lognormal distributions and a close relationship is observed between their scaling parameters,\footnote{The ratio between their mean and their standard deviation or median is nearly the same.} suggesting a likely proportional relationship. This indicates that particular socio-economic groups, on average, purchase vehicles of similar prices.\footnote{Note that the strength of this correlation and its scaling is found to vary across the world, where in some countries, the importance of vehicle purchases relative to other expenditures differ. In particular, in Japan, the car is losing its significance as an identity symbol and expenditure, and the relationship between income and vehicle price is comparatively weakened \cite[see the data in][]{JAMA2013}.}

Based on this evidence, here we infer that car buyers seek to display their social identity, and thus social group characteristics, when purchasing a vehicle, an assertion supported by socio-anthropological theory \citep[the anthropology of consumption, seminal work by][]{Douglas1979}, principles generally accepted in marketing research. Following the reasoning of \cite{Douglas1979}, while all passenger vehicles are built for the purpose of transporting persons, they are marketed for particular social groups at particular prices that match their willingness to pay, and these prices also restrict access to other social groups. The price one is able to pay for a vehicle may be a particularly important channel for the communication of one's social identity, and, as emerges from figure~\ref{fig:UKIncome2012}, appears to be a good indicator of one's disposable income.

The diversity of social groups within a nation therefore strongly influences the diffusion of innovations in the private transport fleet. Formally, the \emph{breadth of consumer diversity determines the elasticities of technology substitution}, an assertion that is best understood using discrete choice theory \cite[for instance in ][]{Ben-Akiva1985}. This assertion is consistent with classical diffusion theory \citep{Rogers2010}, in which the diversity of consumers determine the scaling of the typical  $S$-shaped profile of diffusion. It also follows from evolutionary economics, where technology diversity continually increases to ever better match evolving consumer tastes.

\subsection{Theory for evaluating the effectiveness of policy \label{sect:effectiveness}}

\begin{figure}[t]
		\begin{center}
			\includegraphics[width=1\columnwidth]{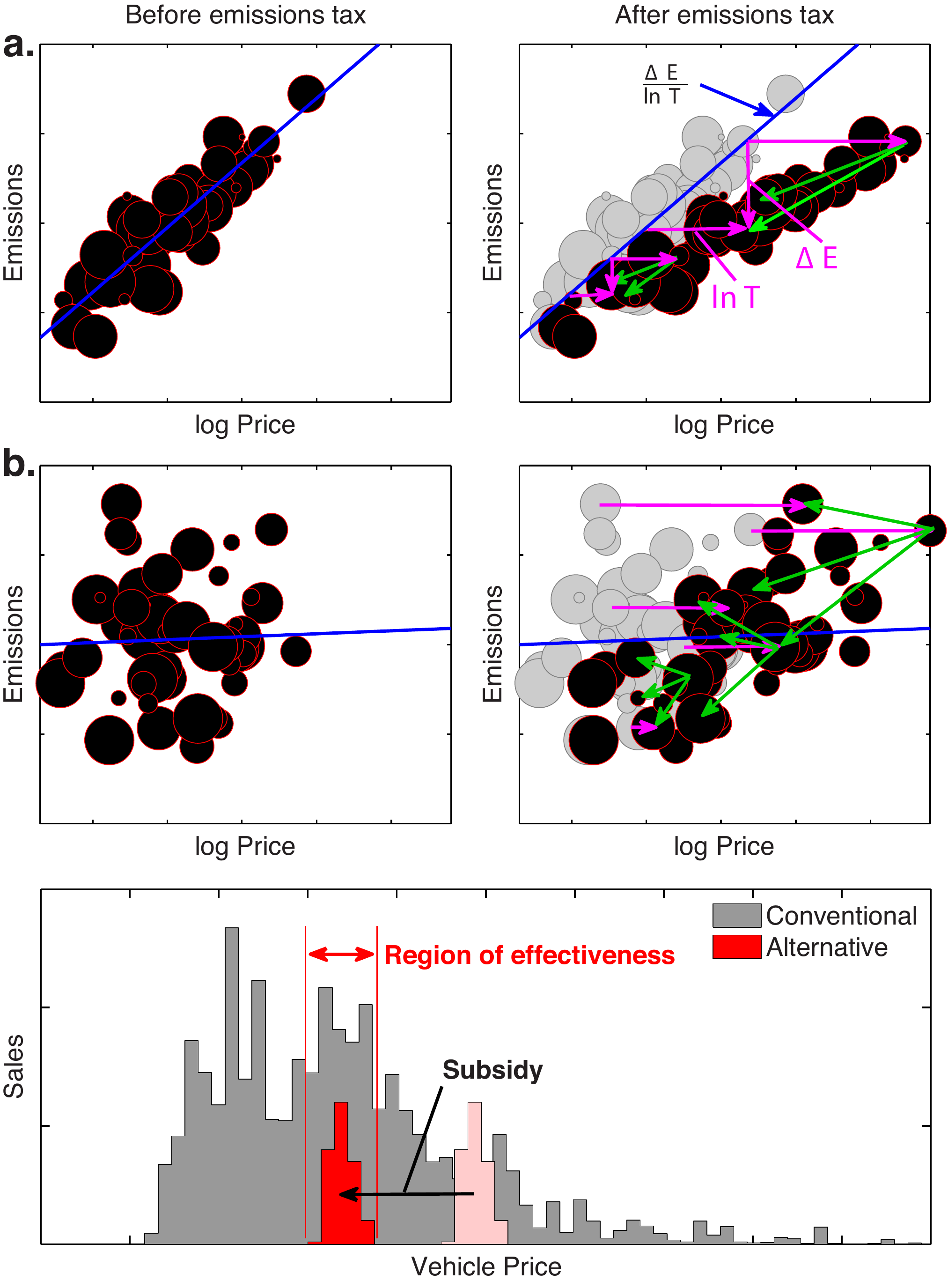}
		\end{center}
	\caption{Schematic representation of the effectiveness of fiscal policies based on emissions in the case where there is (a.) a strong correlation between emissions and vehicle price and where (b.) there is no clear correlation. Blue lines represent linear regressions. With a tax that depends on rated emissions (not necessarily proportional), points in the graph are moved to the left (purple arrows). Possible reactions of consumers changing vehicle choice in order to counteract price changes are suggested with green arrows. In (c.) we illustrate the effectiveness of technological change policies given of a fictitious alternative technology with limited market coverage, where a subsidy is applied.}
	\label{fig:SchematicImpact}
\end{figure}

Emissions reduction policies targeting private vehicle choices have an effectiveness that depends on the structure of local vehicle markets.\footnote{By this we mean markets for new vehicles, since the sale of second hand vehicles does not change fleet emissions, unless they are scrapped before the end of their statistical lifetime, e.g. with a dedicated scrapping policy.} Intuitively, one expects that fuel efficiencies decrease with increasing levels of luxury. As we find here, however, prices do not always scale with emissions, and we describe below the implications to the response of consumers to policy-making.

Vehicles with large engine displacements have in general higher torque and power \citep[for a review of data see][]{WEC2009}. For this enhanced power, more fuel (energy) is used and therefore, for any carbon-based fuel, higher power means higher emissions per kilometre travelled.\footnote{Two aspects of driving behaviour  predominantly underlie a demand for higher power. For the same travel itinerary, (1) higher constant speeds (wind resistance increases faster than linearly with speed, which thus requires more energy per kilometre the faster the vehicle travels). (2) higher acceleration (requires more energy since energy losses are proportionally higher at higher engine revolutions required for higher acceleration).} We explore this relationship below; even though significant amounts of energy may be used in vehicles for other purposes than movement (e.g. air conditioning, electronics), we find that a relationship between emissions and engine size is always measurable, and it is linear.

\setcounter{footnote}{1}

Vehicles with larger engine displacements are generally thought more luxurious and expensive than vehicles with small economical engines. This, however, is not always true and, as we find below, depends on particular vehicle markets. As we show below, when it exists, it is log-linear.\footnote{The log of the price is a linear function of engine size.} Therefore, whether emissions, or the fuel economy, are functionally related in any particular way with vehicle price depends on these two relationships: between emissions and engine size, and between engine size and price. When it exists, it is also of the log-linear type.

The effectiveness of taxes or subsidies applied on vehicle properties such as engine size or emissions corresponds to their ability to generate substitutions between available vehicle models in order to achieve reductions in the emissions of new vehicles. We show here that at the aggregate level, the effectiveness depends on the structure of the market and on whether a relationship between emissions and price exists. If a relationship exists, the effectiveness is closely related to its correlation parameters. If no relationship exists, the aggregate outcome of fiscal policies may or may not turn out to match what was expected in their design, and high levels of uncertainty remains. Given the log-linear structure of the market, an emissions reduction tax on the purchase price that produces a non-negligible incentive across the spectrum of vehicles requires a fee that increases at least proportionally to the emissions rating. When proportional, it is consistent with the `polluter pays' principle. Note, however, that with fiscal policies based on the polluter pays principle, according to our data, the incentive comparatively decreases with increasing car price.\footnote{Car prices increase exponentially for corresponding emissions that increase linearly. At high car prices, emissions taxes become comparatively small.} The effect of taxes on the price of fuels to the choice of vehicles is technically very similar to that of taxes on rated emissions, since the consumer, at purchase time, can only at best estimate future fuel cost using manufacturer rated emissions. Thus we treat them in the same way here. Estimating their effectiveness however requires extra knowledge, or assumptions, on the extent to which consumers evaluate and take consideration of fuel costs over the vehicle's lifetime in their purchasing decisions.

The effectiveness of policy in a diverse market is explained schematically in figure~\ref{fig:SchematicImpact} $a$ and $b$, in which two fictitious vehicle markets are illustrated using one circle per vehicle model, of which the area scales with its number of sales. In \emph{a.}, we have a market where exponentially more expensive vehicles have on average proportionally higher fuel consumption. Applying a fiscal policy based on engine sizes or rated emissions is likely to lead to some emissions reductions if consumers, when replacing a vehicle, attempt to remain within a particular price bracket, as the work of \cite{McShane2012} suggests they would. By seeking a price reduction to compensate for the tax, consumers are forced by available choice to pick in almost every case lower emissions vehicles. In this case the policy effectiveness is well defined, and is equal to the slope of the relationship. Note that this effect also works in reverse: if their income increases, consumers may seek higher price vehicles, which in turn would have higher emissions. Note also that the effectiveness value is independent of the shape of the chosen relation between the tax value and rated emissions (i.e. proportional or not).\footnote{Pigouvian and fuel taxes are proportional to emissions, but other relationships, e.g. exponential, could also be used.}

In \emph{b.}, we have a situation where no correlation exists between emissions and price. In this case, the aggregate impact of a fiscal policy on emissions is ambiguous and could lead to uncertain changes in emissions. This is because there is a wide range of possible fuel economy values that consumers can access while attempting to choose lower price vehicles to compensate for the tax. In this case the policy effectiveness is itself ambiguous. Insight on the effectiveness of policy can thus be obtained from a combination of the strength of the correlation between prices and emissions (the level of confidence that a response to the policy would arise), and the slope of the relationship itself (the strength of the response). 

Meanwhile, emissions can also be influenced using policies supporting changes in engine technologies, such hybrid or electric. Technological change could effectively reduce emissions if alternative engine vehicle models have very low emissions and if they are accessible to most consumers. Currently, however, (see below) most markets do not offer a very wide range of models, which may not have the ability to capture the whole breadth of existing consumer diversity, restricting their diffusion. The effectiveness of subsidies will be determined by whether they help better match new technologies to consumer tastes in market segments where they can be made attractive. This is depicted schematically in panel \emph{c.}, where hypothetical price distributed sales for an alternative technology are shown (in red) along with a typical sales distribution of conventional vehicles. The range of the market that can be expected to be affected by a policy, according to \cite{McShane2012}, is a restricted segment of the whole market. A subsidy policy changes the market segment in which the technology is being marketed. This restricted effectiveness can be altered in the future if manufacturers succeed in broadening market choice in order to cover more social identity groups.

Finally, we may ask, is the analysis of the impact of a tax on the price of fuels conceptually any different to that for a tax on the price of vehicles proportional to emissions? There is considerable controversy on how buyers of new vehicles take consideration of fuel costs when taking a decision \citep[e.g.][]{Busse2013, OECD2010}. In the consumer's perspective, a fuel tax applies later in time than a tax on vehicle prices. There is no clear evidence to support a claim that high emissions vehicles are scrapped more quickly in situations of a high tax on fuels; it mostly changes their trading value in second hand markets, but they likely remain in the fleet \citep{Busse2013}. There is evidence that fuel taxes lead to reductions of car use and associated emissions \citep{Busse2013}, which originates from changes in lifestyles, transport modes, load factors and economic impacts, but this subject is outside of the scope of the present paper, which focuses on consumer choices for new vehicles. Thus for our purposes, for a similar tax value, a fuel tax influences the choice of new vehicles in a similar way as a registration tax proportional to emissions.

The extent to which consumers consider fuel costs when choosing a vehicle can be expressed by how much they discount future fuel costs, and the discount rate expresses their time preference. The literature reports values from around 5-10\% \citep{Busse2013} to 20, even 40\% \citep{OECD2010}, depending how the measurement is done. We show in the SI that fleet averaged lifetime fuel costs are most of the time much lower than fleet averaged car prices. The effectiveness of a tax on fuels, per percent of tax applied, is proportionally lower than a tax on car prices proportional to emissions, by the proportion that fuel costs make in the total cost, which partly depends on the discount rate chosen. Calculations using various rates are given in the SI.

Vehicle markets, policy and consumer choices co-evolve with time \citep[e.g.][]{Geels2006}. It is thus understood that the analysis presented is a present day picture which will evolve as manufacturers, consumers and policies co-evolve. While the market structure will likely change, the short term analysis approach proposed here should remain valid.

\subsection{Calculating the effectiveness of a fiscal policy \label{sect:eqns}}

Given comprehensive market data, the likely effectiveness of a fiscal policy given above can be calculated using a regression. However, more detailed insight may be obtained with a model of distributed choice dynamics. We assume, as in the above, that when imposing a new tax on the purchase of vehicles based on emissions, the response of consumers will be to compensate the tax with a lower vehicle cost. We define a symmetric probability distribution (in log scale) $f(\ln P_i-\ln P_j + \ln T_j, \sigma)$, that determines the approximate region in the $(E,\ln P)$ plane where they search the market. $P_i$ is the price of vehicles $i$ considered before the tax comes into force, $P_j$ is the price of vehicles $j$ that consumers decide to purchase instead once the tax comes into force, however without the tax included, and $r_j = T_j - 1$ is the vehicle dependent tax rate applied on models $j$ based on their rated emissions $E_j$. $\sigma$ is the tolerance of consumers to price differences, as a fraction of price, that we assume within 5 to 20\%. Following \cite{McShane2012}, consumers who would, before the tax, have purchased vehicle model $i$, their probability of choosing model $j$ instead will be proportional to the relative popularity of model $j$, (it's number of sales $N_j$):
\beq
\mathcal{P}_{i \rightarrow j} = {N_j f(\ln P_i - \ln P_j + \ln T_j) \over \sum_k N_k f(\ln  P_i - \ln P_k + \ln T_k)}.
\eeq

Before the tax was applied, $N_i$ consumers per unit time purchased model $i$. After the tax is applied, these consumers will most likely purchase other models instead, while other consumers from another price tier will purchase model $i$. Since there were $N_i$ consumers initially considering model $i$, the number of consumers changing their choice from $i$ to $j$ is thus
\beq
\Delta N_{i \rightarrow j} = N_i N_j {f(\ln  P_i - \ln P_k + \ln T_k) \over \sum_k N_k f(\ln  P_i - \ln P_k + \ln T_k)} \Delta t.
\eeq
Consumers who would have purchased model $j$ also have a non-zero probability of purchasing model $i$,\footnote{Especially if the tax was negative, i.e. a subsidy; this equation must be symmetric under changes of sign of $\ln T_i$.} $\Delta N_{j \rightarrow i}$.  The number of changes of choices between model $i$ and $j$ is
\bea
\Delta N_{i j} &=& \Delta N_{i \rightarrow j} - \Delta N_{j \rightarrow i} \nn\\
&=& N_i N_j \left[ g_{ij}(\ln T_i) - g_{ji}(\ln T_i) \right] \Delta t,
\eea
\beq
g_{ij}(\ln T_i) = {f(\ln  P_i - \ln P_k + \ln T_k) \over \sum_k N_k f(\ln  P_i - \ln P_k + \ln T_k)}.
\eeq
A new set of vehicle sales when the tax applies can be calculated using $\sum_j \Delta N_{ij} + N_i$. Here, we are interested in calculating the average emissions change related to these change of choices. We therefore add up each choice change's impact on total fleet-year emissions:
\beq
\overline{\Delta E} = {1 \over N_{tot}} \sum_{ij} E_j \Delta N_{ij},  \quad N_{tot} = \sum_i N_i,
\label{eq:dE}
\eeq

The cumulative amount of registration tax applied and paid associated to new choices is calculated differently since the tax value is zero before the tax is applied:
\beq
\overline{\ln T} = {1 \over N_{tot}} \sum_{ij}  \left( \Delta N_{ij} + N_i \right) \ln T_j.
\label{eq:dT}
\eeq

$\overline{\ln T}$ is approximately equal to an average tax rate $\overline{r}$ between zero and one. The value obtained from the ratio $\overline{\Delta E} / \overline {\ln T}$ gives the weighted average emissions reductions per unit of average tax paid that result from consumer choice.\footnote{E.g. if $\overline{\Delta E}/ \overline{\ln T}$ = 40 gCO$_2$/km, then for a 10\% average tax one obtains mean emissions reductions of 4~gCO$_2$/km.}

The breadth of the variations between individual choices and the average $\overline{\Delta E}/ \overline{\ln T}$ is obtained using $\delta \overline{\Delta E} = \sqrt{\langle \Delta E^2 \rangle - \langle \Delta E \rangle^2}$, where $\langle \rangle$ indicates a population weighted average.\footnote{We considered that variations on $\overline{\ln T}$ are the same as those on $\overline{\Delta E}$, since the relationship between emissions and tax is exact, and thus this uncertainty is not double-counted.} This may be interpreted as an uncertainty range; however note that it actually corresponds to the standard deviation of a known distribution of individual behaviour, while $\overline{\Delta E}$ is its mean, with significant amounts of variations cancelling in the aggregate. The same calculation is done for engine sizes $\overline{\Delta S}$.

\section{Data and results \label{sect:results}}

\begin{figure*}[p] 
		\begin{center}
			\includegraphics[width=2\columnwidth]{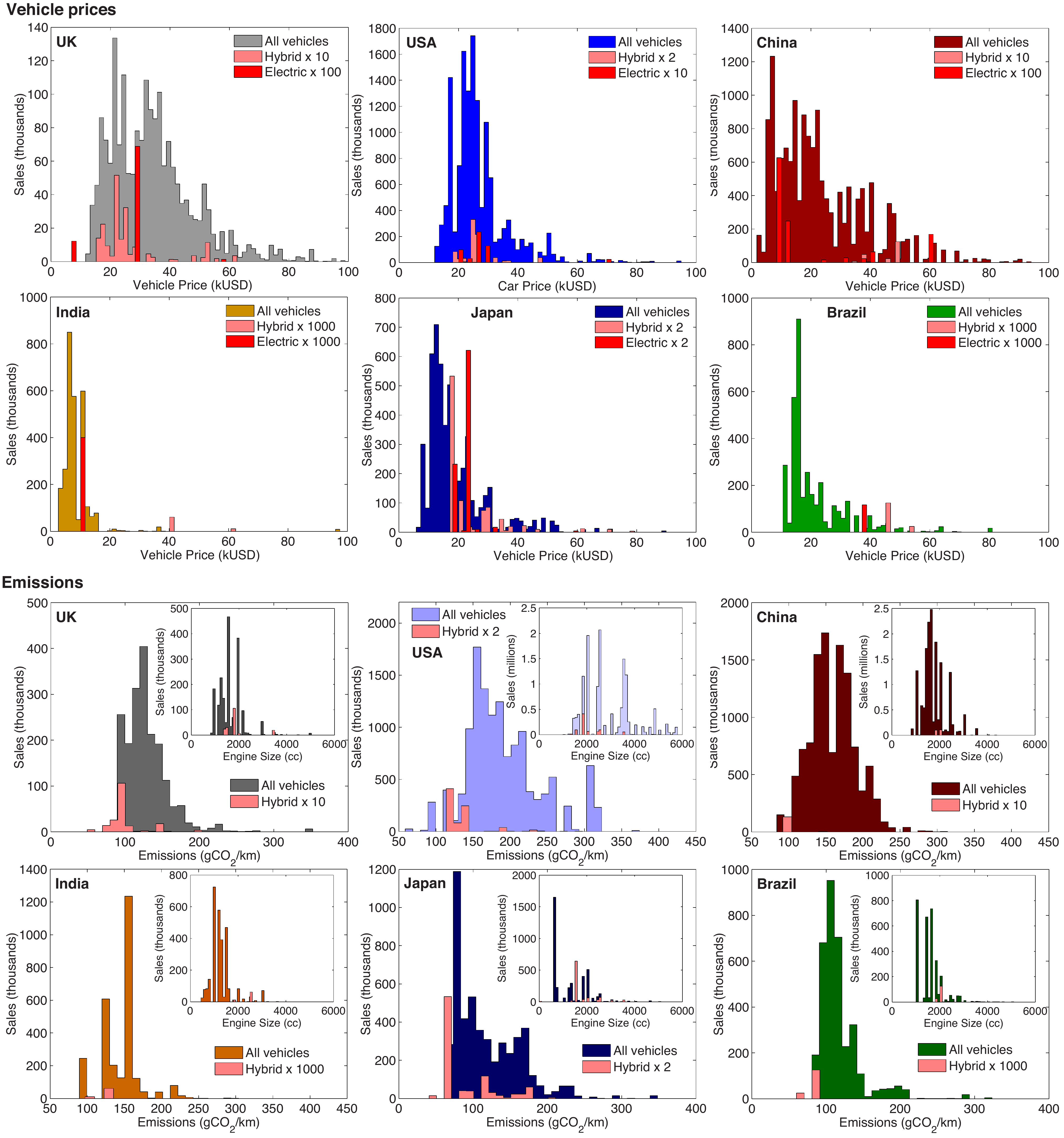}
		\end{center}
	\caption{\emph{Top two rows}: Price distributions of 2012 vehicle sales for six countries on linear price axes with identical price scaling. Price distributions of alternative engine technology vehicles are shown in pink (hybrid) and red (electric), multiplied by the amounts indicated in the legends for legibility. \emph{Bottom two rows}: Emissions distributions (main graphs) and engine size distributions (insets) of 2012 vehicle sales for six countries on linear emissions and engine size axes with identical scaling. Similar distributions of hybrid engine technology vehicles are shown in pink, multiplied by the amounts indicated in the legends for legibility (electric cars have no engine displacements or tailpipe emissions).}
	\label{fig:Figure4}
\end{figure*}

\begin{table*}[t]
	\begin{center}
		\begin{tabular*}{2\columnwidth}{@{\extracolsep{\fill}} l l|r r r r r r }
			\hline
			\multicolumn{4}{ l }{Fleet distribution properties}&&&&\\
			\hline
					&	&	UK&		USA&		China&		India&		Japan&		Brazil\\
			\hline
			\hline
					&Avg&	34285&	25959&	22826&	8674&	18317&	20642\\
			P (USD)	&Med&	31520&	23871&	18970&	6947&	14968&	16425\\
					&Std&	18640&	10570&	16633&	12418&	11526&	13770\\
			\hline
					&Avg&	1576&	3026&	1704&	1220&	1286&	1527\\
			S (cc)	&Med&	1498&	2550&	1596&	1170&	1252&	1558\\
					&Std&	544&		1225&	481&		445&		728&		458\\
			\hline
					&Avg&	123.3&	185.4&	154.1&	140.4&	113.4&	112.2\\
			E (g/km)	&Med&	118.8&	175.6&	152.5&	145.0&	102.2&	105.6\\
					&Std&	30.1&	50.3&	30.6&	26.6&	44.3&	28.5\\
			\hline
			\hline
		\end{tabular*}
	\end{center}
	\caption{Properties of vehicle fleets for six representative countries. This includes average (Avg), median (Med) and standard deviations (Std) of prices, engine sizes and emissions. Price distributions are well described by log-normal, and thus their medians are lower than the means.}
	\label{tab:Avg}
\end{table*}

Sales for new private passenger vehicles were obtained from \cite{Marklines} for all six countries except the UK, for which a more detailed dataset was used for new vehicle registrations from the registration agency \cite{DVLA}. Entries were matched, model by model, to various data sources, all commercial websites, for vehicle price, engine size and rated emissions: \citep{CarPages} (UK), \cite{autousa} (USA), \cite{SohuChina} and \cite{AutoHome} (China), \cite{Zigwheels} and \cite{Carwales} (India), and individual car maker websites. For almost all models matched, we thus obtained their price, rated emissions, engine size and the number sold from the combination of only two data sources by model, enabling to look for correlations. We matched in this way over 4200 models across the six nations.\footnote{2212 in the UK, 470 in the USA, 630 in China, 188 in India, 455 in Japan, 335 in Brazil. All data sets cover all types of private passenger vehicles with 4 wheels (i.e. we excluded buses and motorcycles). Where numbers of models are higher, such as in the UK, more variants of similar models were included. In the UK, the \cite{DVLA} database for new registrations has over 29~000 entries, featuring large numbers of entry variants of similar or identical models. In the UK the matching was restricted to entries with sales of more than 100 units, however for other countries, all Marklines data was used.}

Prices were converted from local currencies to US dollars using \burl{www.XE.com} (June 2014). Rated emissions and engine sizes are those given by the manufacturers. \cite{Marklines} numbers were checked for reliability against total sales given by a number of official data sources, and proved to be reliable. We stress that Marklines data are \emph{comprehensive} in these countries, not samples. Variations of prices for particular models related to optional features were found to remain most of the time within about 5-20\% of the basic model prices, providing a rough basis for the value of~$\sigma$. 

Figure~\ref{fig:Figure4} shows distributions of private passenger vehicle prices on linear price axes for 2012 registrations in our six major economies. Sales of alternative technologies, hybrid and electric cars, are shown in pink and red respectively, scaled up for legibility. All graphs have identical abscissa scaling and binning for comparison. All distributions can be parameterised by log-normal distributions.\footnote{This was determined by both binning the distributions on a log axis and observing distributions that are roughly normal, and by fitting non-linearly log-normal distributions to the data with linear binning, not shown here.}  We provide average and median prices with their standard deviations in Table~\ref{tab:Avg}. The distributions of emissions and engine sizes for the same data are also shown in figure~\ref{fig:Figure4}, with engine sizes in the insets. Average and median emissions and engine sizes, with their standard deviations, are given in Table~\ref{tab:Avg}. 

\begin{figure*}[t] 
		\begin{center}
			\includegraphics[width=2\columnwidth]{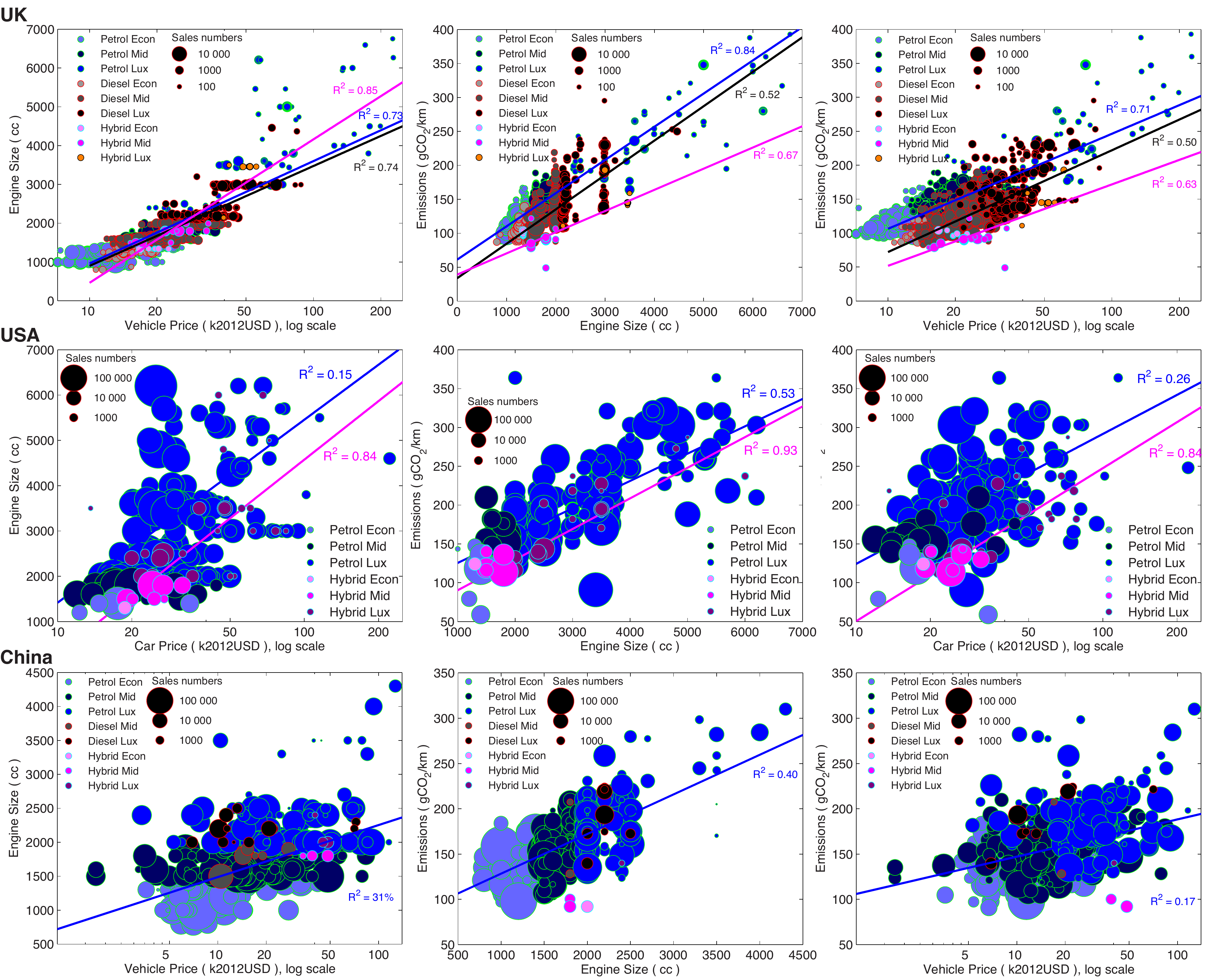}
		\end{center}
	\caption{Population weighted univariate regressions between variables for the UK, USA and China. \emph{Left column}: engine sizes against the logarithm of vehicle prices. \emph{Middle column}: Emissions against engine sizes. \emph{Right column}: Emissions against the logarithm of vehicle prices.}
	\label{fig:Figure5}
\end{figure*}

\begin{figure*}[t]
		\begin{center}
			\includegraphics[width=2\columnwidth]{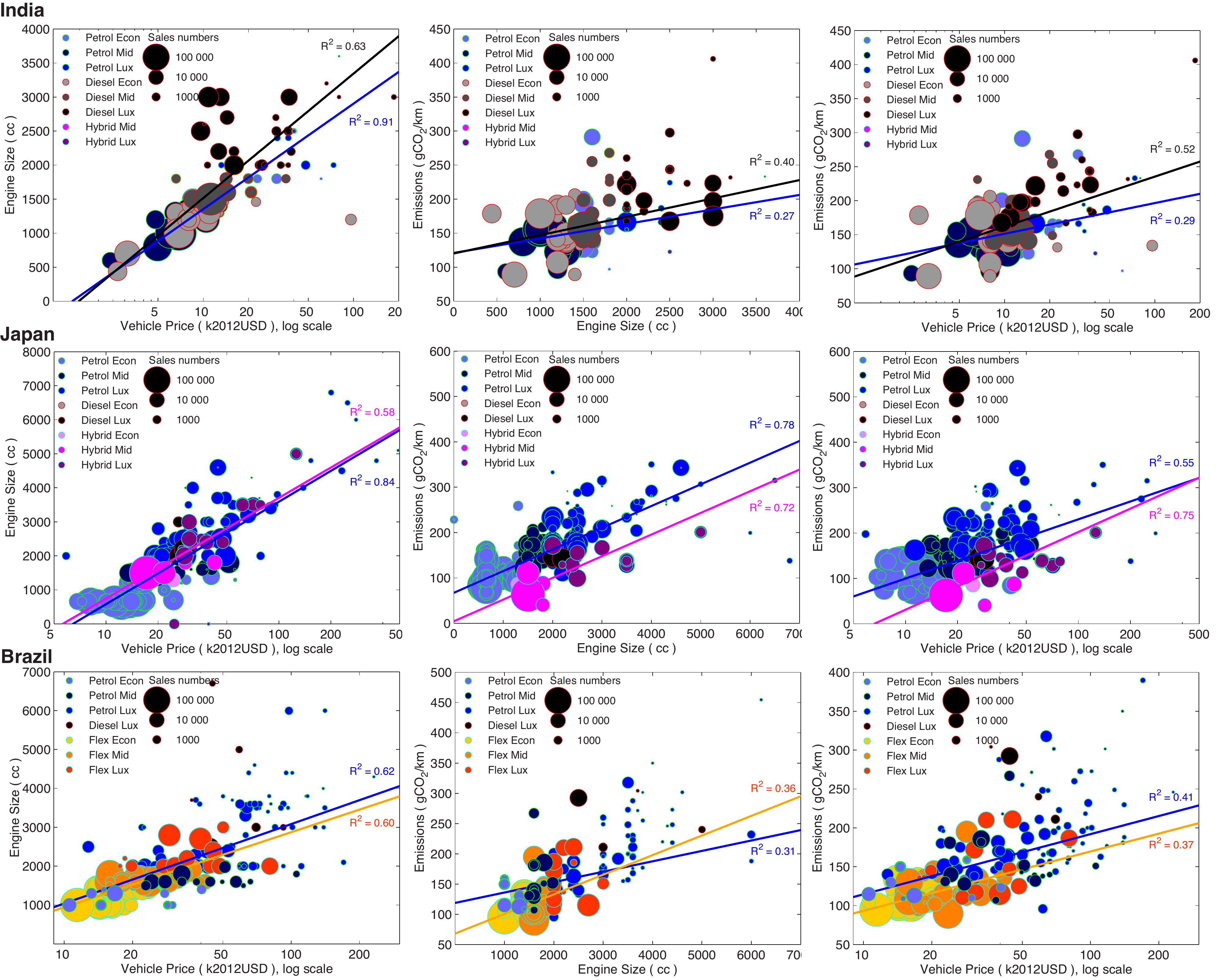}
		\end{center}
	\caption{Population weighted univariate regressions between variables for India, Japan and Brazil. \emph{Left column}: engine sizes against the logarithm of vehicle prices. \emph{Middle column}: Emissions against engine sizes. \emph{Right column}: Emissions against the logarithm of vehicle prices. In the case of Brazil, flex indicates cars that can use either petrol or ethanol.}
	\label{fig:Figure6}
\end{figure*}

\setcounter{footnote}{1}

Scatter plots with linear univariate correlations between engine sizes and the logarithm of vehicle prices are shown in the left hand panels of figures~\ref{fig:Figure5} and \ref{fig:Figure6}. The same is given for emissions and engine sizes, middle panels, and for emissions and the logarithm of prices, in the right hand panels. In these bubble graphs, one circle is shown per model, of which the area is scaled with the root of the number of sales.\footnote{The root was used for illustrative purposes, to increase the visibility of less popular models and to limit the area of circles for popular models. This enabled the same scaling to be used in all graphs.} The same circle size scaling was used in all graphs. Lines are population weighted linear fits of two variables, with coefficients of determination $R^2$ indicated (all parameters given in Table~\ref{tab:Corr}). We did not assume that emissions were correlated with engine sizes and prices simultaneously (the regressions are not multivariate); instead we considered that emissions depend on vehicle prices exclusively \emph{through} the relationship between emissions and engine sizes. Uncertainty values on regressions were obtained using a Monte-Carlo method (details in the SI). 

Table~\ref{tab:Corr} gives correlations parameters and effectiveness values ($\overline{\Delta E} / \overline {\ln T}$) calculated using eqns.~\ref{eq:dE}~and~\ref{eq:dT}, for both types of vehicle tax schemes based either on emissions or engine sizes. The effectiveness values per unit of average tax paid were calculated using three values of consumer tolerance to price changes $\sigma$, with uncertainty, over a range of 42 tax schemes proportional to emissions or engine sizes, with values of between -3kUSD up to 20kUSD at 300~gCO$_2$/km, for taxes on emissions, and between -3 up to 20kUSD/cc at 5000cc for taxes on engine sizes. The negative ranges were included to test whether negative taxes lead to higher emissions, which effectively leads to the same but negative effectiveness values (see details in the SI). This indicates that increases in relative income leads consumers to choose higher emissions vehicles by the same scaling parameter. The effectiveness of a fuel tax, since it is very similar to a registration tax proportional to emissions, is explored in the SI, however since when including fuel costs in consumer decisions, this tax is expressed in terms of a smaller component of the total operating cost, the effectiveness value obtained is proportionally smaller.\footnote{I.e. this only means that a doubling of the price of fuels has a smaller effectiveness than a tax of which the fee results in a doubling of the average price of the vehicles.}

Finally, long term impacts of fiscal policies supporting low carbon engine technologies at low levels of diffusion is difficult to carry out using this short term analysis: it requires modelling diffusion dynamics. Here we find however that in the short term, emissions were not always reduced in the model when subsidising the purchase of hybrid cars since the policy may actually lead consumers to replace low carbon petrol and diesel vehicles by higher emissions hybrids. This depends strongly on the specific details of the policy, which needs more careful study. Meanwhile, subsidising electric cars in all cases decreases emissions, but by amounts orders of magnitude lower than the values of table~\ref{tab:Corr} in which all types of vehicles are involved. 

\section{Discussion \label{sect:Discussion}}

\subsection{Distributions}

We observe that the structures of vehicle markets are widely different across countries. Within developed nations, while the price distribution in the UK is comparable to that of the USA, the distribution in Japan is lower and narrower. For the emerging nations, in India and Brazil the distributions are narrow and similar to that of Japan, while the distribution in China is broad, comparable to that in the UK. This is likely related to income distribution, culture and social dynamics within those societies. These variations do not have any clear relationship to GINI coefficients for these countries or other measures of inequality.\footnote{While a relationship with income distribution apparently exists (as suggested by figure~\ref{fig:UKIncome2012}), the relationship with inequality is likely more complex.}

We also find disparities between distributions of engine sizes across nations, seemingly unrelated to any particular physical or geographical features or constraints. These are unrelated to differences in price distributions. For example, despite similar price distributions in the UK and the USA, the distribution of engine sizes in the USA covers 1000 to 6000cc, while that of the UK is concentrated between 1000 and 3000cc. 

We find similar disparities across countries in the distributions of rated emissions. Cars in the USA have the broadest emissions distribution, while Brazil and India have the narrowest, followed by the UK and Japan. The USA has the highest average and median emissions, while Brazil and Japan have the lowest. The upper end of the emissions distribution in the USA has a value of almost twice that in Brazil or India, and 50\% higher than the UK and Japan. These market structure differences are identified as tied to cultural and behavioural characteristics of consumers, but also tied to their regulatory and tax history (including fuel taxes, which are markedly lower in the USA).

Price distributions of alternative technologies are also very different across countries. The UK is the nation with the widest choice for hybrid cars while it is the USA for electric cars. Japan sees the highest market penetration in absolute numbers for both, however these cover only about the upper half of the price range. The availability and penetration of alternative technologies in emerging nations is very low, and in India and Brazil the price of hybrid vehicles is prohibitively higher than what people spend on vehicles. This means that in the USA and the UK, most consumers can access an alternative vehicle technology in price ranges near what they are willing to pay for a vehicle. Meanwhile this is not the case in emerging countries where the market coverage of these technologies is very patchy. Japan is in an intermediate situation. This limits highly the effectiveness of alternative technology support policies such as subsidies.

The emissions distributions of hybrid vehicles have lower averages than those of ICE vehicles. However their emissions are not always lower than the fleet-year averages, where for example in the USA, many hybrid vehicles have higher emissions than most petrol cars. A similar observation can be made about Japan and the UK. As the availability of technologies in these countries is comparatively high, this feature correlates with the fact that emissions of hybrid vehicles sold to consumers purchasing in large engine size brackets are higher than emissions of any vehicles in lower engine power brackets; however emissions of high power hybrids are still lower than those of the vehicles they most likely replace in the same price or engine size brackets. It suggests that subsidising hybrid vehicles can lead to increased emissions.

\begin{table*}[t]
\tiny
	\begin{center}
		\begin{tabular*}{2\columnwidth}{@{\extracolsep{\fill}} l l| c c c|c c c|c c c }
			\hline
			\multicolumn{11}{ c }{\multirow{2}{*}{Correlation parameters}}\\
			\multicolumn{11}{ c }{}\\
			\hline
			\multicolumn{2}{ l|}{} & \multicolumn{3}{ c|}{Engine size vs log Price} &\multicolumn{3}{ c|}{Emissions vs Engine Size} &\multicolumn{3}{ c }{Emissions vs log Price} \\
				&		&$a^1$		&$b^2$			&$R^2$	&$a^3$		&$b^4$		&$R^2$	&$a^5$		&$b^6$		&$R^2$\\	
			\hline	
			\hline
			UK	&Petrol	&1145	&-9581		&0.73	&48.87	&61.3	&0.84	&60.8	&-453.2	&0.71\\
				&Diesel	&1118	&-9395		&0.74	&50.6	&33.9	&0.52	&65.1	&-527.5	&0.50\\
				&Hybrid	&1605	&-14314		&0.85	&31.2	&39.2	&0.67	&52.0	&-427.3	&0.63\\
				&\bf All	&\bf1089$\pm$76	&\bf-9072$\pm$785		&\bf0.76	&\bf40$\pm$4	&\bf64$\pm$6	&\bf0.56&\bf43$\pm$6	&\bf-314$\pm$57	&\bf0.42\\
			\hline
			USA	&Petrol	&1755	&-14747		&0.15	&35.2	&90.2	&0.53	&72.7	&-545.2	&0.26\\
				&Hybrid	&1868	&-16929		&0.84	&39.5	&50.7	&0.93	&85.6	&-738.0	&0.84\\
				&\bf All	&\bf1751$\pm$333	&\bf -14742$\pm$3357	&\bf 0.13	&\bf 36$\pm$9	&\bf 87$\pm$21&\bf 0.54	&\bf 73$\pm$21	&\bf -545$\pm$207 &\bf 0.23\\
			\hline
			China&Petrol	&333		&-1578		&0.32	&43.8	&84.6	&0.40	&17.9	&-17.6	&0.17\\
				&Diesel	&259		&-623		&0.31	&97.9	&-19.2	&0.31	&36.2	&-148.9	&0.14\\
				&Hybrid	&456		&-2976		&0.05	&23.7	&51.6	&0.40	&-44.6	&573.4	&0.42\\
				&\bf All	&\bf331$\pm$120	&\bf-1560$\pm$1178		&\bf0.32	&\bf43$\pm$11	&\bf84$\pm$21	&\bf 0.40	&\bf18$\pm$7	&\bf-66$\pm$67	&\bf0.17\\
			\hline
			India	&Petrol	&673		&-4847		&0.91	&21.3	&121.0	&0.27	&19.6	&-29.2	&0.29\\
				&Diesel	&794		&-5796		&0.63	&26.9	&120.2	&0.40	&31.9	&-131.8	&0.52\\
				&\bf All	&\bf790$\pm$273	&\bf-5828$\pm$2455		&\bf0.75	&\bf28$\pm$25	&\bf116$\pm$39	&\bf0.40	&\bf30$\pm$24	&\bf-114$\pm$218	&\bf0.45\\
			\hline
			Japan&Petrol	&1308	&-11466		&0.84	&47.8	&67.2	&0.78	&56.7	&-422.4	&0.55\\
				&Hybrid	&1286	&-11098		&0.58	&47.8	&4.5		&0.72	&74.4	&-653.8	&0.75\\
				&\bf All	&\bf1333$\pm$210	&\bf-11693$\pm$2051	&\bf0.84	&\bf41$\pm$14	&\bf67$\pm$14	&\bf0.67	&\bf50$\pm$20	&\bf-349$\pm$/87	&\bf0.49\\
			\hline
			Brazil&Petrol	&887		&-7130		&0.62	&17.2	&118.8	&0.31	&33.6	&-194.6	&0.41\\
				&Flex fuel	&838		&-6777		&0.60	&32.4	&68.1	&0.36	&33.1	&-211.6	&0.37\\
				&\bf All	&\bf964$\pm$187	&\bf-7980$\pm$1875		&\bf0.84	&\bf35$\pm$18	&\bf67$\pm$24	&\bf0.73	&\bf42$\pm$16	&\bf-328$\pm$157	&\bf0.76\\
			\hline
			\hline
		\end{tabular*}
		\begin{tabular*}{2\columnwidth}{@{\extracolsep{\fill}} l|c c c|c c c }
			\multicolumn{7}{ c }{\multirow{2}{*}{Tax effectiveness ${\Delta E \over \ln T}$ and ${\Delta S \over \ln T}$}}\\
			\multicolumn{7}{ c }{}\\
			\hline
			\multicolumn{7}{ c }{Tax on emissions} \\
			\hline
			& \multicolumn{3}{ c|}{Emissions reductions (in gCO$_2$/km)} &\multicolumn{3}{ c }{Engine size reductions (in cc)} \\
			&$\sigma = .05$	&$\sigma = .1$	&$\sigma = .2$	&$\sigma = .05$	&$\sigma = .1$	&$\sigma = .2$	\\	
			\hline
			\hline
			UK 		&$32\pm15$	&$32\pm11$	&$31\pm8$	&$781\pm157$		&$780\pm118$		&$775\pm92$		\\
			USA 		&$93\pm52$	&$82\pm42$	&$64\pm30$	&$2128\pm673$	&$1941\pm552$	&$1465\pm406$	\\
			China 	&$25\pm22$	&$25\pm19$	&$21\pm14$	&$327\pm231$		&$294\pm214$		&$263\pm167$		\\
			India 	&$38\pm57$	&$37\pm48$	&$32\pm36$	&$717\pm389$		&$637\pm331$		&$549\pm261$		\\
			Japan 	&$26\pm39$	&$34\pm28$	&$35\pm18$	&$859\pm384$		&$874\pm278$		&$884\pm222$		\\
			Brazil	&$33\pm86$	&$32\pm69$	&$26\pm32$	&$820\pm902$		&$748\pm746$		&$584\pm381$		\\
			\hline
			\multicolumn{7}{ c }{Tax on engine size}\\
			\hline
			UK 		&$33\pm12$		&$33\pm9$	&$33\pm8$	&$858\pm139$		&$861\pm112$&$870\pm103$\\
			USA 		&$83\pm56$		&$78\pm44$	&$64\pm32$	&$2361\pm745$	&$2156\pm613$&$1682\pm460$\\
			China 	&$25\pm25$		&$23\pm19$	&$21\pm12$	&$265\pm263$		&$276\pm211$&$302\pm148$\\
			India 	&$26\pm93$		&$22\pm67$	&$17\pm45$	&$781\pm664$		&$731\pm475$&$623\pm347$\\
			Japan 	&$26\pm42$		&$36\pm26$	&$38\pm23$	&$1155\pm470$	&$1208\pm318$&$1206\pm302$\\
			Brazil	&$34\pm77$		&$35\pm60$	&$29\pm32$	&$965\pm850$		&$834\pm656$&$658\pm374$\\
			\hline
			\hline
		\end{tabular*}
	\end{center}
	\caption{(\emph{Top table}) Table of correlation parameters, of the form $y=ax + b$, calculated using the data of figure~\ref{fig:Figure5} and \ref{fig:Figure6}. Units are in ($a^1$) (cc = cubic centimetres, L = litres) cc / log P, ($b^2$) cc, ($a^3$) gCO$_2$/km/L, ($b^4$) gCO$_2$/km, ($a^5$) gCO$_2$/km/log P and ($b^6$) gCO$_2$/km. $R^2$ indicates coefficients of determination expressing the strengths of the correlations. Uncertainty values for correlation parameters were obtained using Monte-Carlo analysis, described in the SI. (\emph{Bottom table}) Calculated values for the effectiveness of taxes on either rated emission or engine sizes are given, with their aggregate impacts on both rated emissions and engine sizes for new potential fleet-years. These are calculated using eqns.~\ref{eq:dE}~and~\ref{eq:dT}, i.e. emissions or engine size reductions expected for 100\% tax applied (see calculation details with examples in the SI). Uncertainty ranges correspond to the standard deviations of variations in individual consumer choices with respect to the means. They should not be interpreted strictly as uncertainty values on the means, since significant amounts of variations in consumer choices cancel out in the aggregate, even when their variations are significant.}
	\label{tab:Corr}
\end{table*}

\subsection{Correlations and policy effectiveness}

Correlations between emissions and engine sizes exist in all countries, and this is due to this relationship being mostly an engineering one (figures~\ref{fig:Figure5} and \ref{fig:Figure6}, middle panels, and Table~\ref{tab:Corr}, middle columns), where larger engines are more powerful, use more energy (fuel) and therefore produce higher emissions per distance driven. The parameters of these correlations are similar but differ across countries, and this may be related to either or both: (1) additional features coming with higher power vehicles that also use energy, of which the purchase differ between countries, or (2) different levels of technology sophistication in vehicles leading to varying levels of fuel efficiency at similar engine sizes.

However, correlations do not always exist between prices and engine sizes (or power, left panels). Their strength (the $R^2$ parameters), when they exist, vary considerably between countries. In particular, while a very clear log-linear relationship exists in the UK, India, Japan and Brazil, the relationships are very weak in the USA and China. This indicates a clear hierarchy between engine size and prices in vehicle markets in the UK, India, Japan and Brazil, but none in the USA or China. It thus emerges that in China and the USA, the size of the engine is not a major determinant of the price in manufacturer marketing decisions, while it is elsewhere.

Depending whether relationships exist between prices and engine sizes, the existing relationships between emissions and engine sizes bridges to possible relationships between emissions and prices, although with weaker correlations than with engine sizes. Thus, wherever relationships exist between engine sizes and prices, they exist between emissions and prices, and conversely where they don't exist. Thus very weak relationships exist between emissions and prices in the USA and China, while they are stronger in other countries. 

The scaling parameters of these relationships (Table~\ref{tab:Corr}) indicate the likely consumer response to fiscal policy in these countries, with a confidence measure given by $R^2$. However, we consider that more reliable values, along with measures of variations in consumer behaviour, are given by using the average effectiveness of policy per average unit tax paid determined using eqns.~\ref{eq:dE}~and~\ref{eq:dT}. According to these, one expects a high response to policy of 0.6 to 0.9~gCO$_2$/km per percent of average tax in the USA (60-90~gCO$_2$/km for 100\% tax), indicating a very high abatement potential with moderate tax values. Meanwhile in all other countries, the fleet-year-average effectiveness lies in the area of 0.3-0.4~gCO$_2$/km per percent of average tax. These values agree with those obtained with regressions. It has been suggested that indirect emissions taxes based on engine sizes are less efficient at reducing emissions than direct taxes \citep{He2011}. From the consumer choice perspective, this is not supported by our data, where little difference is observed in aggregate between the two tax schemes. Therefore, this assertion is likely to be only true from the supplier perspective, where manufacturers are likely to attempt to circumvent engine size-based taxes by designing vehicles with smaller engines for the same amount of power (e.g. with turbo-chargers).

Negative tax values also lead to increases in emissions by the same factors, suggesting that relative consumer income increases lead to increasing emissions. This can explain current trends of increase in engine sizes in the UK,\footnote{Our data, not shown here.} Ireland \citep{Gallachoir2009} and Germany \cite{Zachariadis2013}. Finally, the effectiveness of taxes are orders of magnitude higher in the short term than those obtained when policy only supports the diffusion of alternative engine technologies with subsidies. 

\section{Conclusion and policy implications}

The data and analysis presented in this paper has clear policy implications. We have provided qualitative and quantitative tools that can help understand and determine the likely outcome of chosen policies targeting consumer vehicle choice for emissions reductions. We find that it is possible to calculate average values for the effectiveness of fiscal policies using market data, and the results differ by country analysed. Our data show a high diversity of consumer choices, tied to socio-economic differences, different in every country, with repercussions on fleet-year emissions and prospects for reductions. 

The effectiveness of fiscal policies for incentivising consumer choices towards lower emissions vehicles of any type, according to this work, is of around 0.3-0.4~gCO$_2$/km per percent of a proportional emissions tax applied to vehicle sale prices, in all countries except in the USA where it is of 0.6-0.9~gCO$_2$/km per percent of tax, 2 to 3 times higher, with higher uncertainty. Taxes based on either emissions or engine sizes have similar impacts. However, using on their own policies to support the diffusion of new engine technologies, given our model of consumer choice, has an effectiveness orders of magnitude lower than tax policies that apply to all types of vehicles simultaneously. Fuel taxes, while reducing transport use, have a similar effect on vehicle choices; however since they only apply to a minor component of total costs, they result in a lower impact. These values, however, apply only in the short term. In the long term, the choice of consumers will influence manufacturer marketing strategy, which will alter the range of vehicles offered to consumers, which will in turn again alter consumer choices, and policy will need to be re-assessed over time. The complex co-evolutionary process between consumer choices, manufacturer strategy and policy-making is a subject that requires further research.

We conclude that both the stringency and the structure of policies are important. Fiscal policies based on emissions are essential to reduce fleet-year emissions across the range of consumer groups; policies supporting changes in engine types are not likely highly effective in the short term, and this is due to a patchy market coverage by hybrid and electric vehicles. Meanwhile, we show that fiscal policies following the `polluter pays' principle generate an incentive that decreases with consumer income. If, in order to completely decarbonise private transport (e.g. the UK's 2050 target), policy strategy is to transform the market towards a dominance of alternative technologies, comprehensive technology demand pull and supply push policies must be used in close collaboration with manufacturers, which may yield positive results in the long term. However, both emissions-based fiscal policies and technology policies must be used simultaneously in order to reduce emissions in both the short and the long term.

\section*{Acknowledgements}

The authors wish to thank H. Pollitt and Miyoshi Hiroaki for highly insightful comments. We thank the Energy Systems Seminar participants for lively discussions on the subject. We thank two anonymous referees for their supportive comments. We acknowledge our respective funders, the Three Guineas Trust (A. Lam) and the UK Engineering and Physical Sciences Research Council (EPSRC), fellowship no EP/K007254/1 (J.-F. Mercure).

\section*{References}
\bibliographystyle{elsarticle-harv}
\bibliography{CamRefs}

\newpage
\onecolumn
\setcounter{section}{0}
\setcounter{figure}{0}
\renewcommand{\theequation}{S.\arabic{equation}}
\renewcommand{\thefigure}{S.\arabic{figure}}
\renewcommand{\thetable}{S.\arabic{table}}
\section*{Supplementary Information}

\section{Details of the calculation in the base case}

We provide example calculations of the effectiveness of policy given the dataset for six vehicle year-fleets of the main paper. We create various fictitious emissions taxing schemes and apply them to all vehicle prices according to their emissions. We assume that from year to year, choices within social groups do not change significantly, i.e. that without policy, choices next year would be the same as those observed in our dataset (e.g. the same lognormal distributions of prices). We furthermore assume that choices do change with policy such that consumers seek other vehicle models that cost approximately what they were initially intending to spend, i.e. they search in the vicinity of the market segment their social group is accustomed to (e.g. economic, mid-range or luxury, and sub-segments within that) in a price region that remains constant. Here we exclude fuel costs, which are treated in the next section.

Since they are not likely to find a vehicle at exactly the same price as they were initially planning to spend (i.e. our vehicle market is \emph{discrete}, not continuous), they may have to accept a certain price difference, and for this we assume that they have a tolerance of $\sigma$. The introduction of $\sigma$ enables to further relax assumptions about choices of social groups by making choices broader, including consumers keeping the same purchase plan and paying the full tax. We initially assign to $\sigma$ a flexible value of 0.1, i.e. 10\% of the price. We subsequently vary this value as a sensitivity analysis using 5\% and 20\%. Note that since this price comparison is carried out in log space, what we are assuming is that the \emph{order of magnitude} of the price remains the same before and after the tax within $\sigma$\footnote{E.g. in this scheme, it would be difficult to find a credible tax regime that would convince buyers of \$100k vehicles to purchase \$10k vehicles.}, and thus this is not a highly constraining assumption.

We follow the equations of Appendix A and present here graphs of results for these calculations. In principle, any tax scheme can be tested with these equations; in this particular case we test a case where the tax value, i.e. the fee, increases proportionally with rated vehicle emissions, consistent with a typical Pigouvian tax on emissions.\footnote{As argued in the main paper, in such a scheme, the incentive for reductions to high price bracket consumers declines with price, since the tax increases linearly with emissions, while as we know from the data, the price rises exponentially with emissions. This, however, does not change the method of the calculation or the result, which is expressed in average emissions reductions per unit of average tax paid.} We then vary the magnitude of the scaling of this tax. New prices (after tax) in terms of old prices (before tax), when expressed in log form (here $r_i = T_i - 1 \simeq \ln T_i$ is a tax \emph{rate}, but in principle need not be a fraction of the vehicle price), are thus
\beq
\log P_j \simeq \log P_i + \log T_i.
\label{eq:tax}
\eeq
We assume for simplicity that the relative probability of choice between two vehicles at slightly different prices away from the centre value of the choice price follows a normal distribution in log scaling (i.e. a lognormal probability), i.e.
\beq
f(\ln P_i - \ln P_j + \ln T_j) = {1 \over \sqrt{2 \pi} \sigma} \exp \left({-{(\ln P_i - \ln P_j + \ln T_j)^2 \over 2 \sigma^2}} \right).
\eeq
This essentially means that vehicle options for which eq.~\ref{eq:tax} is not approximately true within $\sigma$ are highly improbable (undesirable to the consumer). The top left panel of figure~\ref{fig:TaxRates} shows 42 scenarios of tax based on emissions. The tax is proportional to emissions rated by the manufacturer using a standard driving cycle, measured in tons of carbon dioxide per kilometre. 

\begin{figure}[p]
		\begin{center}
			\includegraphics[width=1\columnwidth]{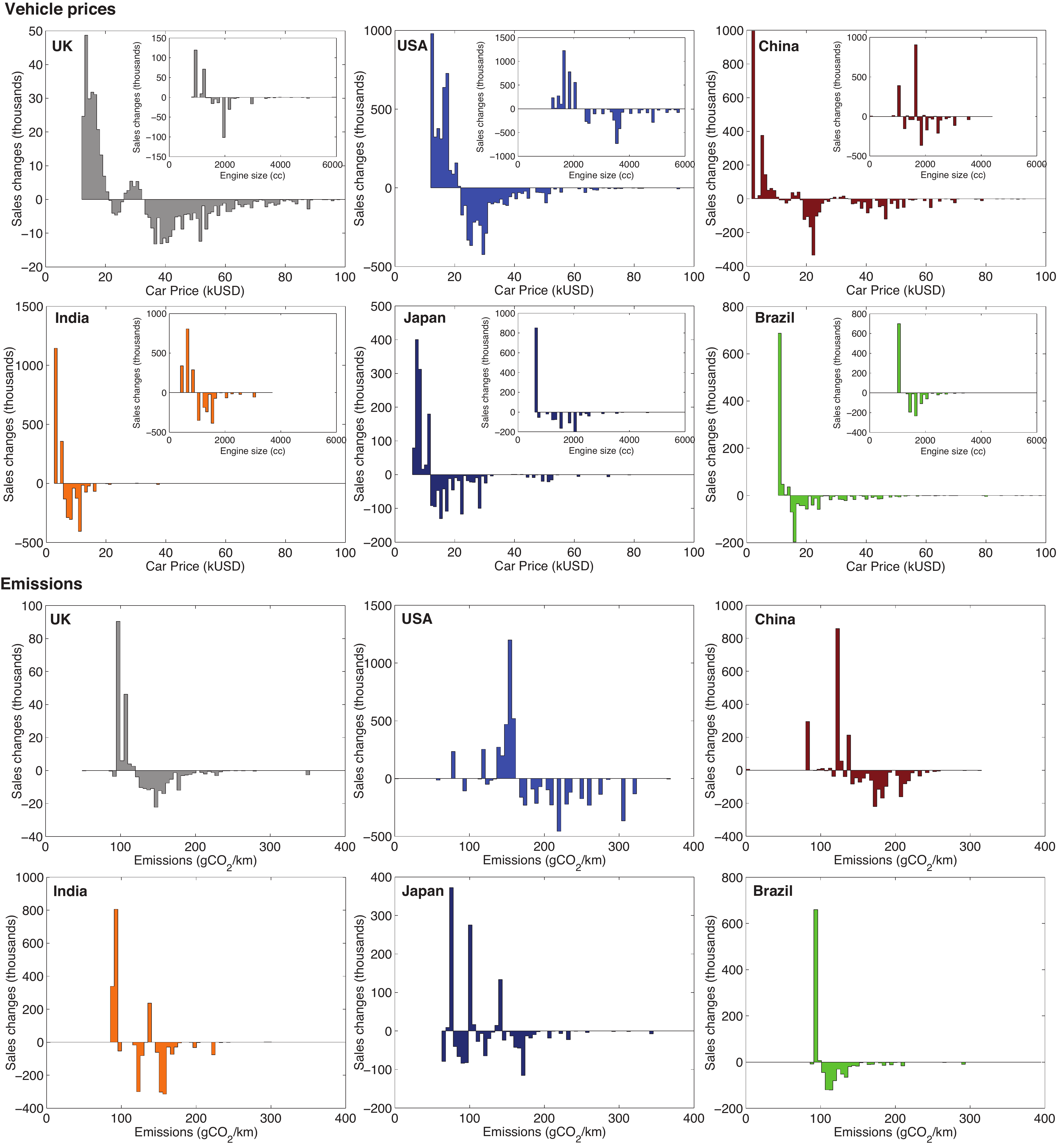}
		\end{center}
	\caption{Changes in the fleet-year due to changes in choices related to a tax proportional to emissions, in this case of a value of 10kUSD at 300~gCO$_2$/km. (\emph{Top six panels}) Changes expressed in terms of prices excluding the tax. Insets show changes in terms of engine sizes. (\emph{Bottom six panels}) Changes in terms of rated emissions.}
	\label{fig:PChoice}
\end{figure}

According to eq.~\ref{eq:tax}, consumers intending to purchase a vehicle before the tax at price $\ln P_i$ will, after the tax, seek a vehicle in price band $\ln P_j \pm \sigma$. Thus while carrying out the calculation, we may explore these choices by charting using distributions of consumer choices in price and emissions space. This is given in figure~\ref{fig:PChoice}, where we chose one of the taxing schemes above where the value of the registration tax at 300gCO$_2$/km is \$10k. Bars below zero indicate price/emissions ranges where sales have decreased, and above zero where sales have increased. These histograms thus represent the changes on the fleet-year that result from the tax according to our model.

New choice of vehicles lead in most cases to choices of vehicles with lower emissions ratings. This would strictly always be the case if the correlations between $E$ and $\ln P$ had a coefficient of determination $R^2 = 1$. Due to scatter in the data, this does not happen, and thus there always exist cases where the choice of vehicles at new prices before tax lower than old prices ($P_j < P_i$) leads to new emissions higher than old emissions ($E_j > E_i$), although this might not be the majority of cases. This means that there are instances where consumers are able to save money (i.e compensate the tax) with choices that increase emissions. 

\begin{figure}[p] 
		\begin{center}
			\includegraphics[width=.9\columnwidth]{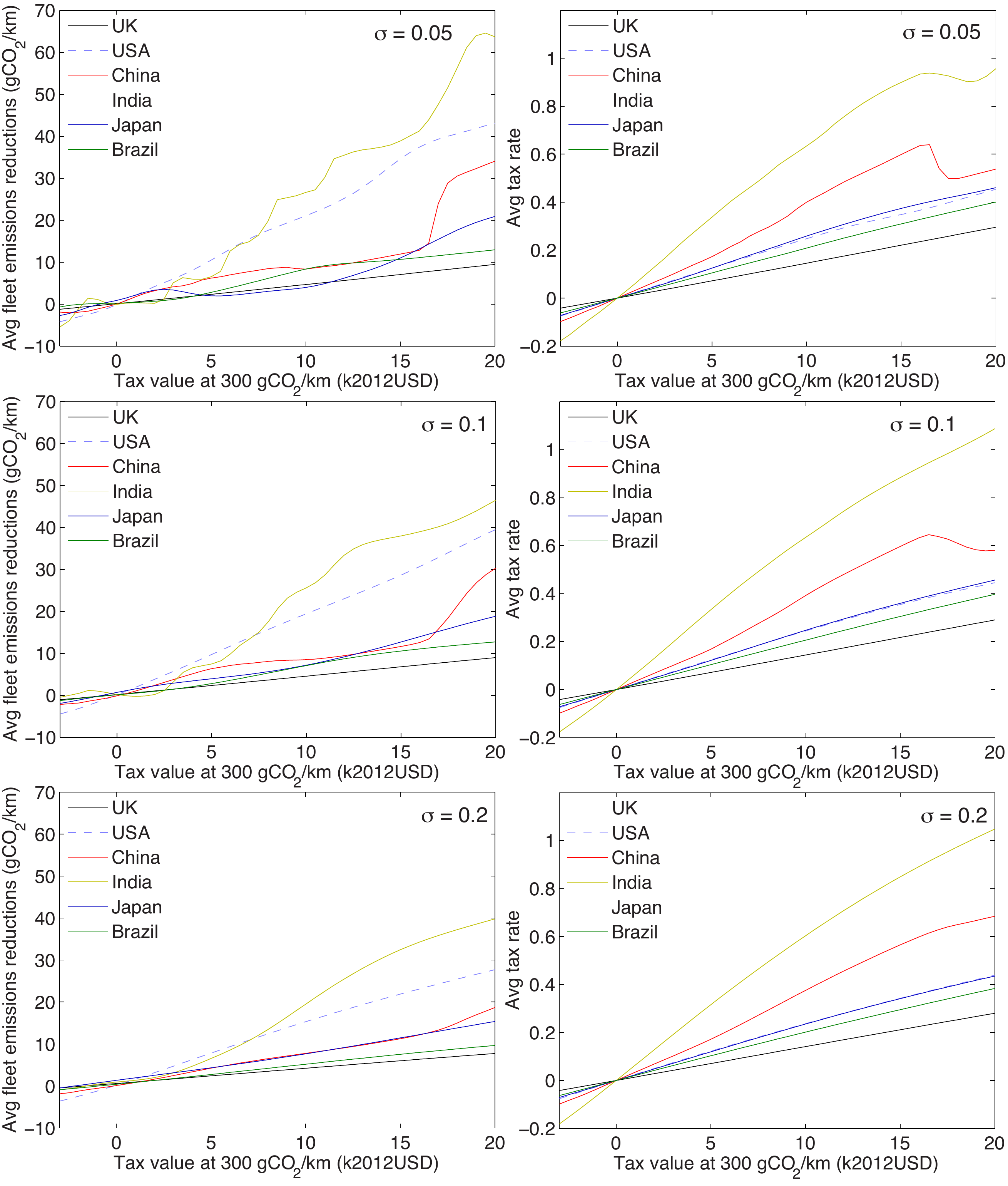}
		\end{center}
	\caption{(\emph{Left column}) Emissions reductions averaged across the new fleet-years as functions of the tax applied (expressed as the value at 300~gCO$_2$/km) for three values of the tolerance $\sigma = 0.05, 0.1, 0.2$. (\emph{Right column}) Tax rate paid by consumers averaged across the fleet for the same tolerances.}
	\label{fig:dEavg_dTavg}
\end{figure}

\begin{figure}[t] 
		\begin{center}
			\includegraphics[width=.9\columnwidth]{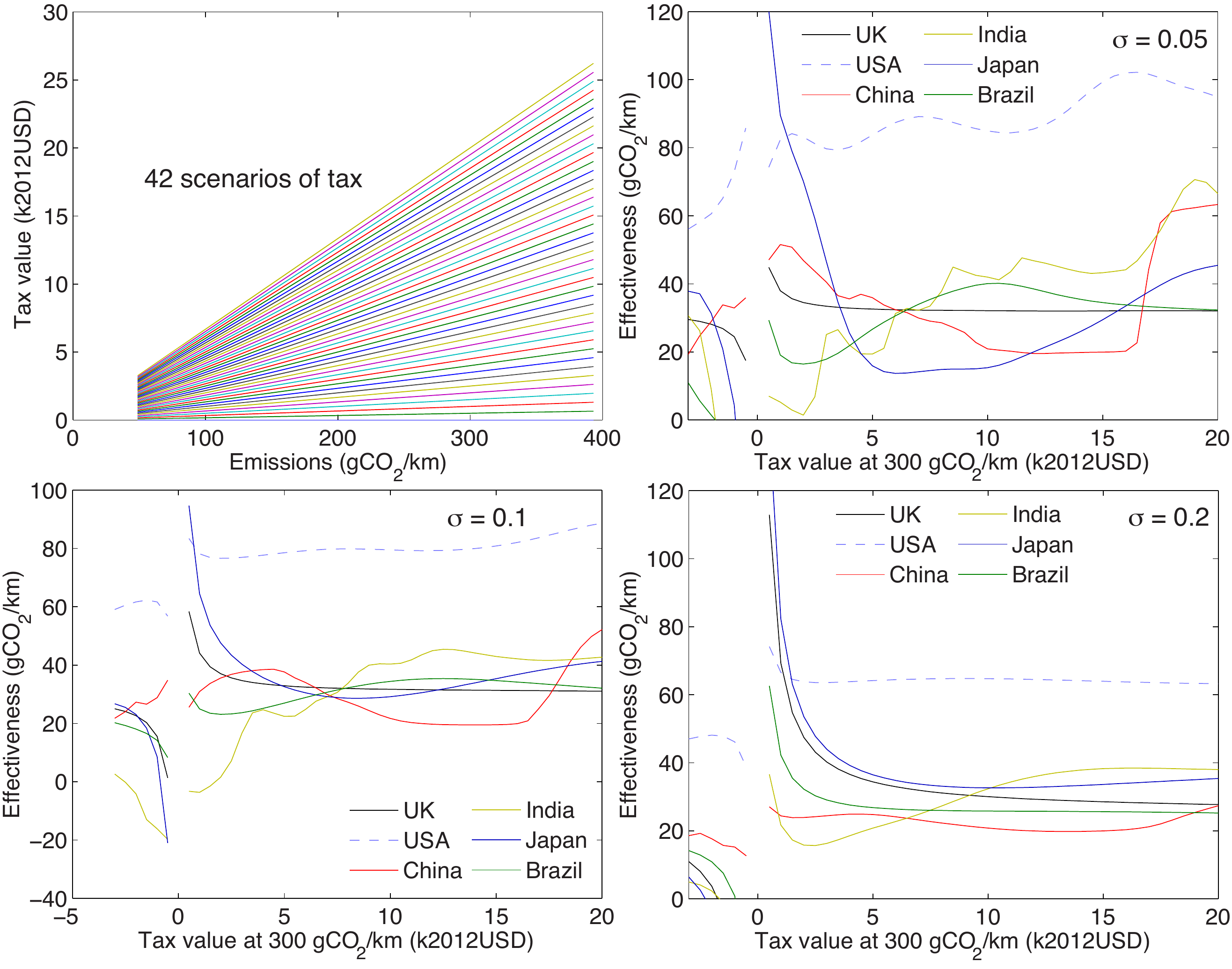}
		\end{center}
	\caption{(\emph{Top left}) Value of 42 different registration tax scenarios. (\emph{Other panels}) Calculation of the effectiveness of the tax for different values of the tolerance $\sigma$ of consumers to price differences.}
	\label{fig:TaxRates}
\end{figure}

Resulting emissions reductions averaged across the fleet-years, for 42 tax scenarios, as functions of the tax value are given in figure~\ref{fig:dEavg_dTavg}, along with the tax paid by consumers also averaged across the fleet-years, for three values of the tolerance $\sigma$. With this one can see how aggregate emissions reductions occur, as a result of a very high number of underlying changes of choice from vehicle model to vehicle model. Meanwhile the average tax paid also results from new choices that accommodate consumers. Here, the $x$-axis indicates the absolute value of the tax at 300~gCO$_2$/km. The value of the tax for other emissions bands can be calculated as a linear proportion, or taken from the first panel of fig.~\ref{fig:TaxRates}. 

Average emissions reductions per average unit of tax rate paid are shown in fig.~\ref{fig:TaxRates} for three different values of the tolerance for price differences $\sigma$. These graphs correspond to the ratio between the traces of figure~\ref{fig:dEavg_dTavg} for each country. Without noise in the correlations and with perfect market coverage in the dataset, we expect in principle (from the argument of section~\ref{sect:effectiveness}) that this quantity should be a constant equal to the slope of the $E, \ln P$ relationship, the effectiveness of the fiscal policy.

We observe different behaviour of this quantity in different countries, and changes that relate to the value of the tolerance. When using a low tolerance value, results appear noisy in countries where the data is more aggregated. For example, our dataset features a choice between 2200 vehicle models in the UK, therefore a near to continuous price distribution of choices, which enables to use a low tolerance (consumers more often find models available near the to the amount they were hoping to spend). In this case, the effectiveness of the tax (emissions reductions per unit tax) is nearly constant for any tax value, except at low values, where the calculation approaches a value of zero divided by zero and small amplitude noise leads to a divergence instead of the expected null effectiveness at zero tax. We stress that the UK dataset is formed by an ensemble of vehicle models each available in a good number of variants at slightly different prices. While the DVLA dataset (see main paper) featured around 30k model variants, our price data featured over 8000 model variants, and our data matching between these was limited to 2200 evenly distributed models due to lack of information over the exact correspondence of the others. We stress that the UK dataset only features more variants of the same models than the other datasets, which are more aggregated under more generic model categories.

In cases of lower quality datasets, as it is the case for India, the data forces a concentration of choices to a small granular set of highly popular vehicle models, and this choice is not entirely constant as function of the value of the tax (lumpy jumps occur at particular tax values). It impossible to distinguish in our data between a lower diversity in the Indian market or a lower resolution of information in the data collection. We observe however that increasing the tolerance damps this effect, as one intuitively expects. Since we know that in reality, most vehicle models are likely available at a wide number of possible price values depending on particular choices of vehicle options (as with the UK dataset), with better data the calculation would most likely reach a stable value in every case.

Table~\ref{tab:CorrS} presents values of the average effectiveness of policy per unit tax paid, as obtained from correlations calculations of the main paper, and from the calculation carried out here, in this case for three values of $\sigma$ of 0.05, 0.1 and 0.2. As $\sigma$ increases, it is observed that the uncertainty generally tends to decrease. However the boundary values of 0.2 and 0.05 can perhaps be considered unlikely themselves. All values are  consistent with the slopes of the correlations. We consider that the results of the calculation given here are more accurate than the values of the correlations. We thus observe that on average, the effectiveness of policy per average percentage point of tax applied on emissions is of around 48~$\pm$~34~gCO$_2$/km per unit tax across countries. This means that at a tax of 100\%, a reduction of emissions of 48~gCO$_2$/km is expected to take place. However the USA has a much higher response value that the other countries. Excluding america, the value becomes 30~$\pm$~30~gCO$_2$/km.

\begin{figure}[t]
	\begin{center}
		\includegraphics[width=1\columnwidth]{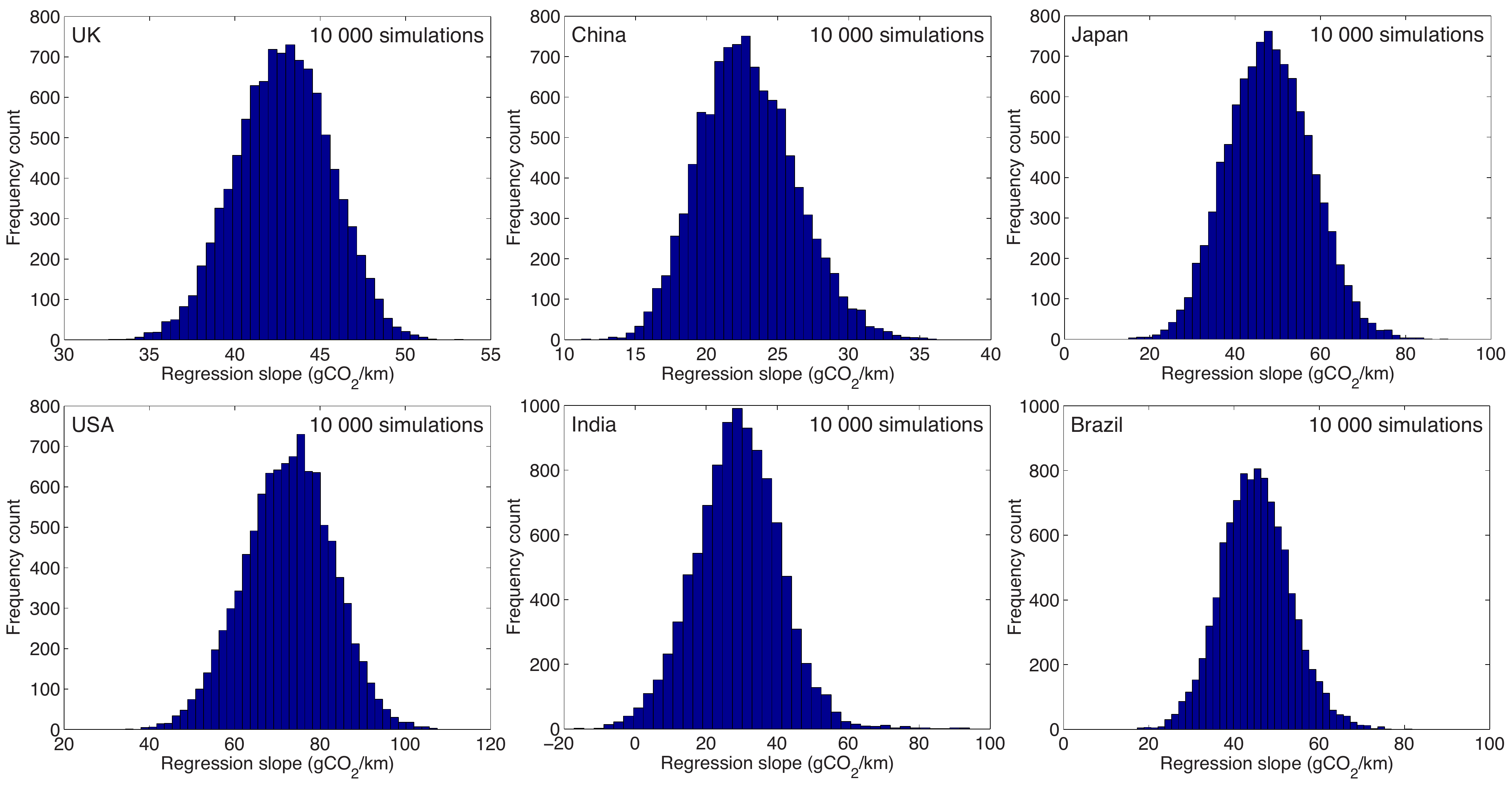}
	\end{center}
	\caption{Monte Carlo analysis of parameter fit uncertainty for the slope of the $E, \ln P$ relationships in the data.}
	\label{fig:MC}
\end{figure}

\begin{table}[h!]
\footnotesize
	\begin{center}
		\begin{tabular*}{.7\columnwidth}{@{\extracolsep{\fill}} l| c c r r c c c c }
			\hline
			\multicolumn{8}{ c }{Effectiveness values $\Delta E / \ln T$} \\
			\hline
					&Correlation	& $R^2$	&Models	&Sales	&$\sigma = .05$	&$\sigma = .1$	&$\sigma = .2$	& Avg.\\	
			\hline
			UK		&\bf 43$\pm$6		&.42	&2207	&1.9M	&33$\pm$15	&\bf 32$\pm$11		&31$\pm$8	&32$\pm$11\\
			USA		&\bf 73$\pm$21	&.23	&188		&13.0M	&93$\pm$52	&\bf 82$\pm$42 	&64$\pm$30	&78$\pm$41\\
			China	&\bf 18$\pm$7 		&.17	&464		&11.1M	&25$\pm$22	&\bf 25$\pm$18 	&21$\pm$14	&28$\pm$17\\
			India		&\bf 30$\pm$24	&.45	&81		&2.7M	&38$\pm$57	&\bf 37$\pm$48 	&32$\pm$36	&37$\pm$47\\
			Japan	&\bf 50$\pm$20	&.49	&236		&4.9M	&26$\pm$39	&\bf 34$\pm$28 	&35$\pm$18	&31$\pm$28\\
			Brazil	&\bf 42$\pm$16	&.76	&185		&3.4M	&33$\pm$86	&\bf 32$\pm$69 	&26$\pm$32	&31$\pm$62\\
			\hline
			\hline
		\end{tabular*}
	\end{center}
	\caption{Table of parameters and averages of calculation results for the effectiveness of policy. The correlation parameters (in units of emissions reductions per unit tax, gCO$_2$/km) and $R^2$ parameters are the same as those given in the main paper (table~2). The numbers of models correspond to subsets of the data for which values were available for the number of sales, their price, emissions ratings and engine size simultaneously. The last three columns are values of the effectiveness of policy, averaged over all tax scenarios above \$5k at 300gCO$_2$/km, in units of average emissions reductions per unit of average tax paid, calculated as described above using three values of the tolerance parameter $\sigma$. Uncertainty values correspond to two standard deviations.}
	\label{tab:CorrS}
\end{table}

These values can be compared to those of the correlations calculated in the main paper. Correlations generate scaling parameters which may be sensitive to the choice of data points within the dataset. In particular, correlations calculated using the whole dataset may not always generate the same values as correlations carried out using subsets of the data. In order to test this, simple Monte Carlo techniques can be used. Here, we calculated the correlation parameters 10 000 times using each time half the number of data existing points chosen randomly. From the outcomes, we carried out statistics in order to determine the uncertainty over the scaling parameters. The frequency count of the outcomes are given in figure~\ref{fig:MC} for the case of the $E, \ln P$ relationship, from which we derived means and standard deviations. The resulting uncertainty is given in table~\ref{tab:CorrS} and in table~2 of the main text. The same was carried out for the other two relationships, with results indicated in table~2 of the main text.

\section{Including fuel costs to the calculation}

Here we carry out the same calculation, with the addition of fuel costs. Fuel costs, from the consumer's viewpoint at the time of choosing a vehicle model, can only be estimated very approximately by using the rated emissions, or fuel economy, of vehicles considered. However fuel costs arise during the lifetime of the vehicle, and their estimation, in the modeller's view of how consumers take decisions, requires us to know to which extent they take them into consideration. The level at which consumers take these into consideration can also be expressed using a discount rate, which represents the value of future expenditures at the time of decision. There is no clear consensus in the literature on how consumers in the vehicle market consider fuel costs. This however, as we will see here, depends highly on social groups, since the fraction of total costs made by future fuel costs varies significantly, which complicates the analysis.

\begin{figure}[t]
	\begin{center}
		\includegraphics[width=.5\columnwidth]{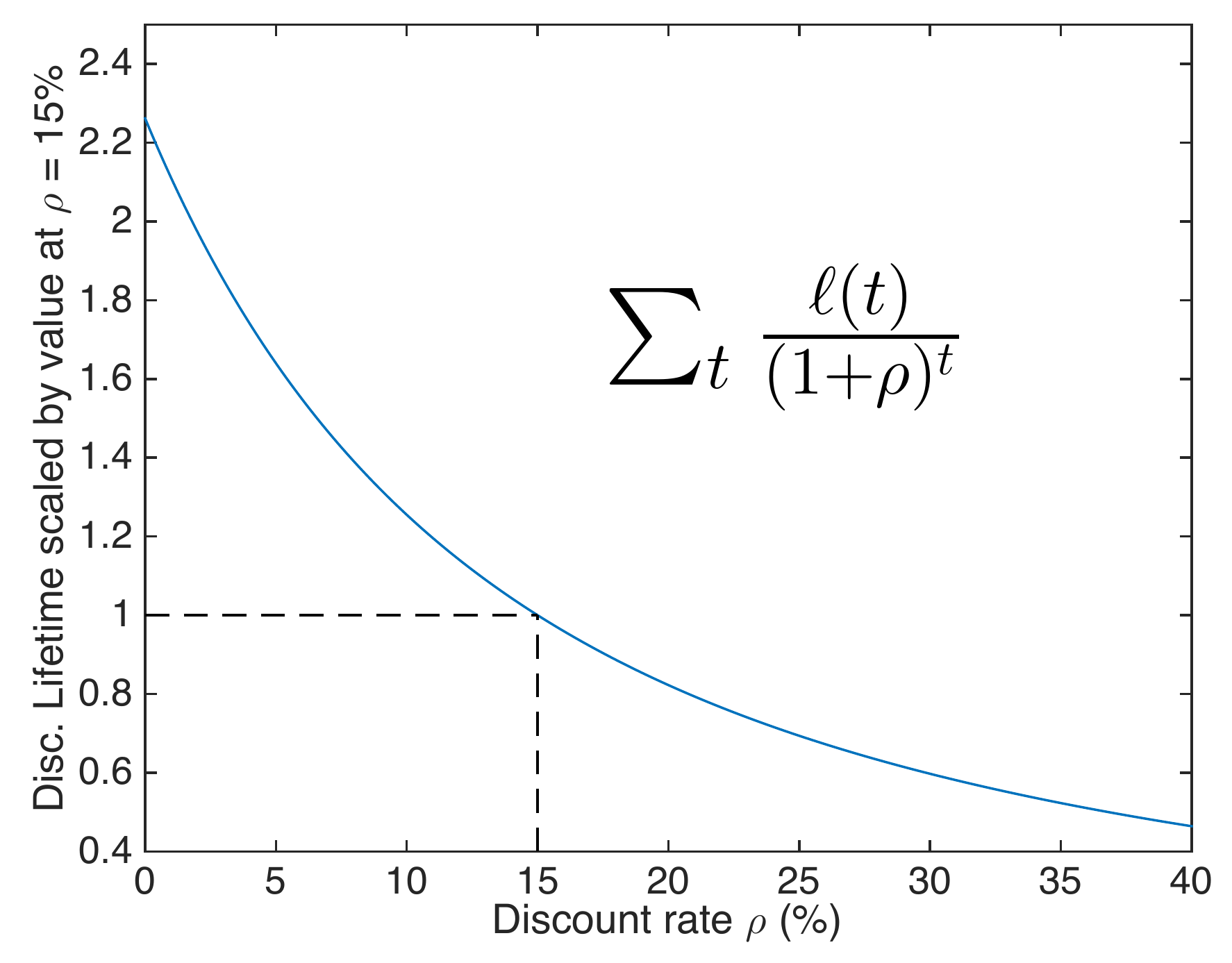}
	\end{center}
	\caption{Factor to rescale results in table \ref{tab:Fuel} when changing the base discount rate of 15\%.}
	\label{fig:Disc}
\end{figure}

Vehicles have a survival probability, of which consumers are aware at decision time. Vehicles are widely known to survive, without accidents, between 10 and 20 years, and therefore it is not realistic to expect consumers to take account of fuel costs much after of the life expectancy of 12 years. We denote the survival function $\ell_i(a)$, the probability of a vehicle surviving to age $a$. If we assume that the consumer effectively uses a discount rate $\rho$ to determine the present value of future fuel costs when choosing a vehicle, then the lifetime fuel cost is
\beq
C^F_i = \sum_a {\ell_i(a) d_i(a) P^F(a) E_i \beta_i \over (1 + \rho)^a},
\label{eq:disc}
\eeq
where $d_i$ is the average yearly expected distance travelled with the vehicle considered, $P^F$ are expected future fuel prices (in \$/L) and $\beta$ a constant that scales the rated emissions value $E_i$, in gCO$_2$/km, into a fuel economy in L/km. Using survival data from the \cite{DVLASurvey, DVLA}, average distances driven from \cite{Euromonitor}, fuel prices from the \cite{WorldBank}, the value of this expression, in terms of the discount rate, is given in figure~\ref{fig:Disc}, while all values are given in table~\ref{tab:Fuel} using a discount rate of 15\%. Here we used constant 2012 values for fuel prices (i.e. as if no changes were expected) and for distances travelled (i.e. same distances are expected to be travelled every year).

\begin{table}[t]
\footnotesize
	\begin{center}
		\begin{tabular*}{.7\columnwidth}{@{\extracolsep{\fill}} l|l| r r r r r r }
			\hline
			\multicolumn{8}{ c }{Lifetime fuel costs, using $\rho$ = 15\% and a life expectancy of 12 years.} \\
			\hline
						&Units	& UK 	&USA		&China	&India		&Japan			&Brazil		\\	
			\hline
			Fuel prices		&\$/L				&2.17	&0.97	&1.37	&1.25	&2		&1.39	\\
			Fuel prices		&\$/tCO$_2$		&957.59	&428.05	&604.56	&551.61	&882.57	&613.39	\\
			Dist. Travelled		&km/y			&12104	&12485	&14145	&23674	&12961 	&28000	\\
			Yearly fuel cost		&\$-km/y-gCO$_2$ 	&11.59	&5.34	&8.55	&13.06	&11.44 	&17.17	\\
		$\sum \ell(a)/(1+\rho)^a$	&No Units			&5.16	&5.16	&5.16	&5.16	&5.16 	&51.6	\\
			Avg  Efficiency		&gCO$_2$/km	 	&123.3	&185.4	&154.1	&140.4	&113.4 	&112.2	\\
			Lifetime fuel cost	&\$				&7374	&5112	&6800	&9460	&6693 	&9943	\\
			Avg veh. price		&\$				&34285	&25959	&22826	&8674	&18317 	&20642	\\
			Fuel \% of total cost	&\%				&18		&16		&23		&52		&27		&33		\\
			Veh. \% of total cost	&\%				&82		&84		&77		&48		&73		&67		\\
			Fuel \% of veh. cost	&\%				&22		&20		&30		&109		&37		&48		\\
			Veh. \% of Fuel cost	&\%				&465		&508		&336		&92		&274		&208		\\
			\hline
			\hline
		\end{tabular*}
	\end{center}
	\caption{Parameters of and results from the calculation of fuel costs at a discount rate of 15\%}
	\label{tab:Fuel}
\end{table}

Using a total cost including discounted fuel costs in the calculation of the effectiveness of taxes opens up the possibility of examining possible taxes on fuels as well as on vehicle registration (road taxes that are paid every year amount to the same calculation as for fuel taxes unless we discount in smaller increments than yearly, which would yield a minute difference). However, after applying the discounting of eq.~\ref{eq:disc}, it does not require any differences in calculation than what is given in the main text (section~\ref{sect:eqns}), using only a different scaling. In other words, if, as in the UK, fuel costs account, on average, for 18\% of total costs (table~\ref{tab:Fuel}), assuming that consumers value them with 15\% discounting, then the effectiveness of an average tax of 1\% on fuels is 22\% of the effectiveness of an average tax of 1\% on car prices. Table~\ref{tab:EffnessFuel} gives re-calculated effectiveness values when including fuel costs, using the same method as in the main paper.

\begin{table}[h]
\tiny
	\begin{center}
		\begin{tabular*}{1\columnwidth}{@{\extracolsep{\fill}} l|c c c|c c c }
			\multicolumn{7}{ c }{\multirow{2}{*}{Tax effectiveness ${\Delta E \over \ln T}$ and ${\Delta S \over \ln T}$}}\\
			\multicolumn{7}{ c }{}\\
			\hline
			\multicolumn{7}{ c }{Emissions tax on registration} \\
			\hline
			& \multicolumn{3}{ c|}{Emissions reductions (in gCO$_2$/km)} &\multicolumn{3}{ c }{Engine size reductions (in cc)} \\
			&$\sigma = .05$	&$\sigma = .1$	&$\sigma = .2$	&$\sigma = .05$	&$\sigma = .1$	&$\sigma = .2$	\\	
			\hline
			\hline
			UK 		&$35\pm12$	&$34\pm9$	&$32\pm7$	&$786\pm134$		&$785\pm105$		&$762\pm91$		\\
			USA 		&$87\pm49$	&$81\pm36$	&$67\pm27$	&$1980\pm625$	&$1793\pm486$	&$1351\pm391$	\\
			China 	&$43\pm18$	&$36\pm14$	&$27\pm12$	&$496\pm181$		&$424\pm159$		&$364\pm136$		\\
			India 	&$56\pm63$	&$40\pm41$	&$21\pm21$	&$401\pm613$		&$388\pm376$		&$288\pm195$		\\
			Japan 	&$36\pm35$	&$40\pm25$	&$41\pm17$	&$955\pm374$		&$885\pm255$		&$819\pm197$		\\
			Brazil	&$35\pm64$	&$32\pm45$	&$28\pm22$	&$778\pm708$		&$654\pm503$		&$488\pm252$		\\
			\hline
			\multicolumn{7}{ c }{Tax on the price of fuels}\\
			\hline
			UK 		&$7\pm3$			&$7\pm2$		&$7\pm2$		&$169\pm29$		&$169\pm20$		&$164\pm20$\\
			USA 		&$17\pm10$		&$16\pm7$	&$13\pm5$	&$390\pm123$		&$353\pm77$		&$266\pm77$\\
			China 	&$13\pm5$		&$11\pm4$	&$8\pm4$		&$148\pm54$		&$126\pm40$		&$108\pm40$\\
			India 	&$62\pm69$		&$44\pm45$	&$23\pm23$	&$438\pm669$		&$424\pm212$		&$314\pm212$\\
			Japan 	&$13\pm13$		&$15\pm9$	&$15\pm6$	&$349\pm137$		&$323\pm72$		&$299\pm72$\\
			Brazil	&$17\pm31$		&$15\pm22$	&$13\pm10$	&$375\pm341$		&$315\pm122$		&$235\pm122$\\
			\hline
			\hline
		\end{tabular*}

	\end{center}
	\caption{Tax effectiveness values calculated using the equations of section~\ref{sect:eqns} while integrating the cost of fuels of equation~\ref{eq:disc}, where the tax in $\Delta E / \ln T$ refers to a tax on the vehicle price or on the price of fuels. When using a consumer discount rate of 15\%, this gives specific proportions of the total cost in the vehicle and fuel prices, as given in table~\ref{tab:Fuel}, but this can be easily changed using values in figure~\ref{fig:Disc} for other discount rates. Values of effectiveness of a tax on fuels is proportionally lower by its contribution to the total cost, and similarly for registration taxes on the vehicle price. For instance, if we were to use a discount rate of 2\% (40\%), the values of effectiveness of a tax on fuel prices would be $\simeq$2 times larger ($\simeq$ 0.5 times smaller), while the effect of a registration on the vehicle price would be scaled by $1-\alpha \beta$, where $\alpha$ is the value taken from figure~\ref{fig:Disc} and $\beta$ is the contribution of fuel prices (row 9 of table~\ref{tab:Fuel}).}
	\label{tab:EffnessFuel}
\end{table}

When comparing table~\ref{tab:EffnessFuel} with table~\ref{tab:Corr} of the main paper, we find that a tax on the vehicle price based on emissions is not changed significantly if we include fuel costs in the calculation of agents for their vehicle choice. Effectively, it increases the total cost but the tax only applies to a fraction of that cost. This is the case for all countries where fuel costs make a relatively small portion of all costs, i.e. all countries excluding China and India. In the latter, since fuel costs make a higher proportion of total costs, the effectiveness of a tax on emissions is increased if we assume that consumers consider the fuel costs in their choice with a discount rate of 15\%. These values can be re-interpreted to other discount rates by using the proportional scaling factor given in figure~\ref{fig:Disc} (i.e. the value can be $\simeq 2$ times larger at very low discount rates, or $\simeq 2$ times smaller for very high discount rates).

We also find the effectiveness of that a tax on fuels at reducing emissions is generally much lower than a registration tax of the same level, in percent of the price of the quantity taxed. In other words, a tax that on averages doubles the price of vehicles is more effective than a tax that doubles the price of fuels by the factors given in the last row of table~\ref{tab:Fuel}. This is due to the relative fractions of lifetime fuel costs relative to the car price. We stress, however, that there is no clear evidence of how consumers take consideration of fuel costs, and that the effectiveness of taxes on fuels may be lower. Variations between countries in their respective taxes on fuels do, however, likely explain some of the differences observed in the distributions of the fuel economy. They have helped increase with time the efficiency of the fleets in countries where they have been higher (e.g. UK, Japan), where in some cases they are of over 100\%. However, registration taxes based on emissions could in principle be even more effective.

\end{document}